\documentclass[%
reprint,
superscriptaddress,
groupedaddress,
showpacs,preprintnumbers,
aps,
prb,
]{revtex4-2}

\usepackage{amsmath}
\usepackage[]{graphicx}
\usepackage{makeidx}
\usepackage{verbatim}
\usepackage[colorlinks,urlcolor=blue,linkcolor=blue,citecolor=blue]{hyperref}
\usepackage{dcolumn}
\usepackage{bm}
\usepackage{multirow}
\usepackage{float}
\usepackage{booktabs}
\usepackage{tabularx}
\usepackage{makecell}
\usepackage{siunitx}
\usepackage{array,mathtools,amssymb,booktabs,longtable}
\usepackage[caption=false,subrefformat=parens,labelformat=parens]{subfig}
\usepackage{adjustbox}
\usepackage{amsfonts}
\usepackage{ragged2e}
\usepackage[dvipsnames]{xcolor}
\usepackage{hyperref}

\usepackage{braket}
\usepackage{ulem}

\newcolumntype{C}{>{$}c<{$}}
\AtBeginDocument
{
	\heavyrulewidth=.08em
	\lightrulewidth=.05em
	\cmidrulewidth=.03em
	\belowrulesep=.65ex
	\belowbottomsep=0pt
	\aboverulesep=.4ex
	\abovetopsep=0pt
	\cmidrulesep=\doublerulesep
	\cmidrulekern=.5em
	\defaultaddspace=.5em
}

\graphicspath{{figures/}}                       

\newcommand{\Gammabar}{\overline{ \Gamma}}
\newcommand{\Zbar}{\overline{ \mathrm{Z}}}
\newcommand{\Mbar}{\overline{ \mathrm{M}}}

\begin{document}
	\title{One-dimensional Dirac modes in the core of a pentagonal
		topological crystalline insulator nanowire}
	\author{Saeed Samadi}
	\thanks{These authors contributed equally to this work.}
	\author{Rafa\l{} Rechci\'{n}ski}
	\thanks{These authors contributed equally to this work.}
	\author{Marta A. Chabowska}
	\author{Ryszard Buczko}
	\email[]{buczko@ifpan.edu.pl}
	\affiliation{Institute of Physics, Polish Academy of Sciences, Aleja Lotnik\'{o}w 32/46, 02668 Warsaw, Poland}

	\date{\today}
	\begin{abstract}
		We investigate the electronic band topology of recently fabricated pentagonal IV--VI semiconductor nanowires, which contain five radial $\{111\}$ twin planes meeting at the nanowire axis. Tight-binding calculations show that, when the bulk band structure is inverted and the twin planes in the nanowire are cationic, the spectrum contains two spatially separated helical Dirac crossings near $\Gammabar$: one localized at the core and the other at the outer surface. When the twin planes are anionic, the corresponding spectra remain gapped. The crossings originate from the hybridization of five helical channels associated with the twin-plane edges, whose odd number leaves one Kramers pair near the nanowire axis and the other at the outer boundary. Realistic multiorbital calculations for $\mathrm{Pb}_{0.4}\mathrm{Sn}_{0.6}\mathrm{Te}$ predict well-developed core and surface modes at nanowire thicknesses of approximately 50~nm and above. These results establish pentagonal SnTe-class nanowires as an experimentally accessible realization of spatially separated helical channels bound to the axial defect and the outer boundary.
	\end{abstract}
	\maketitle
	
	\section{Introduction} \label{sec:intro} 
	
	IV--VI compounds are narrow-gap semiconductors with unusual electronic properties and established applications in infrared detection and photovoltaics~\cite{Wei1997,Sukhovatkin1542,sargent2009infrared,Semonin1530}. Among them, SnTe has an inverted band gap and realizes a topological crystalline insulator (TCI) phase, in which crystalline symmetries protect Dirac-like metallic surface states~\cite{hsieh2012topological,Ando_TCI}. The same phase occurs in substitutional alloys such as $\mathrm{Pb}_{1-x}\mathrm{Sn}_{x}(\mathrm{Te},\mathrm{Se})$, where the band inversion and hence topological phase can be controlled through composition~\cite{tanaka2012experimental,xu2012observation,dziawa2012topological,safaei2013}.
	
	Quantum confinement and crystalline defects give rise to several low-dimensional manifestations of band topology in SnTe-class materials. Two-dimensional TCI and quantum spin Hall phases have been predicted in SnTe thin films~\cite{Liu_2014spin,Ozawa_2014,safaei2015quantum,Liu_2015}, while atomic-height steps on the (001) surface have been found to bind robust gapless modes~\cite{sessi2016robust,rechcinski2018topological,Iaia2019,Brzezicki2019}. One-dimensional channels can also occur as hinge states of higher-order topological insulators (HOTIs)~\cite{benalcazar2017quantized,benalcazar2017electric,langbehn2017reflection}. In particular, suitably distorted SnTe was proposed to realize a helical HOTI phase, with conducting modes at hinges between mirror-related surfaces~\cite{schindler2018higher}.
	
	SnTe-class nanowires (NWs) provide a natural setting for such phenomena and have also attracted broader experimental and theoretical interest.  Defect-free SnTe NWs with a [001] growth axis have been grown by molecular beam epitaxy (MBE) on graphene, while narrower NWs have been synthesized using alloy nanoparticles~\cite{Sadowski_2018,Liu_2021NWs}. Theoretical studies of such cubic SnTe NWs have investigated their corner and hinge modes and their potential for realizing Majorana bound states~\cite{nguyen2022NW}, while first-principles calculations have revealed a strong thickness dependence of their electronic and topological phases, including a transition from trivial insulating behavior toward the TCI regime~\cite{Hussain2024CubicNW}. More recently, [110]-oriented SnTe NWs were shown to support mirror-protected phases and, under suitable superconducting and symmetry-breaking perturbations, Majorana end modes~\cite{Kawala2025SnTe110}. Experimental progress also includes the in-plane growth of $\mathrm{Pb}_{1-x}\mathrm{Sn}_{x}\mathrm{Te}$ NWs and the electrical characterization of SnTe nanoflakes and nanowires~\cite{Schellingerhout2023PbSnTeNW,Mientjes2025SnTeNFNW}.
	
	Recently, pentagonal $\mathrm{Pb}_{1-x}\mathrm{Sn}_{x}\mathrm{Te}$ NWs were fabricated by MBE along the $[011]$ crystallographic direction, providing an experimentally accessible departure from conventional cubic NW geometries~\cite{hussain2024pentagonal}. These fivefold-twinned structures consist of five rocksalt domains separated by radial $\{111\}$ twin planes (TPs), which meet along a one-dimensional partial disclination at the NW axis. Such structures are stabilized by the competition between reduced surface energy and the elastic cost associated with the twin boundaries and central disclination~\cite{Hofmeister2004Fivefold,BalettoFerrando2005}. Unlike in conventional multiply twinned nanostructures, however, the TPs in SnTe-class materials may affect not only the atomic structure and strain, but also the topology of the electronic bands. In particular, when the bulk band structure is inverted, cationic and anionic $(111)$ TPs have distinct topological character~\cite{Samadi2022_TP}, whereas no such distinction arises in topologically trivial PbTe.
	
	Ref.~\cite{hussain2024pentagonal} investigated the structural and electronic properties of pentagonal PbTe and SnTe NWs using density-functional theory (DFT), semiclassical modeling, and a simplified tight-binding (TB) approach. For both TP sublattice types, the calculations revealed a twofold-degenerate "core-chain" band localized near the NW axis and connecting the valence and conduction bands. Its presence in both PbTe and SnTe shows that its existence does not depend on bulk band inversion. Moreover, the calculations of Ref.~\cite{hussain2024pentagonal} were restricted to ultrathin NWs, for which states localized near the core and outer surface cannot become spatially decoupled.
	
	In this work, we determine the electronic structure and topological classification of pentagonal SnTe-class NWs in the experimentally relevant large-radius regime. We find that NWs with cationic TPs support two helical Dirac crossings near $\Gammabar$, one localized near the NW core and the other distributed over the outer surface, whereas the corresponding spectra of NWs with anionic TPs remain gapped. In contrast to the band reported in Ref.~\cite{hussain2024pentagonal}, which is tied to a high-symmetry stoichiometric core configuration and disappears under unconstrained structural relaxation, these modes require an inverted bulk band structure and persist for different microscopic realizations of the NW axis, including occupied and hollow cores. The results admit a natural microscopic interpretation in terms of five helical channels associated with the cationic TP edges, whose hybridization leaves a single Kramers pair at the core and a corresponding pair at the outer surface. The resulting spatially separated channels are analogous to the opposite edges of a quantum spin Hall ribbon. We establish their origin and symmetry protection using the minimal $p^3$ TB model for SnTe together with a low-energy $k\cdot p$ theory, and confirm their emergence in the material-specific $sp^3d^5$ model of $\mathrm{Pb}_{0.4}\mathrm{Sn}_{0.6}\mathrm{Te}$. These results establish pentagonal SnTe-class NWs as an experimentally accessible realization of spatially separated helical channels localized at the NW core and outer boundary.
	
	\section{Geometry of pentagonal IV--VI nanowires}
	\label{sec:IV-VI}
	
\begin{figure}[!htbp]
		\centering
		\includegraphics[width=0.49\textwidth]{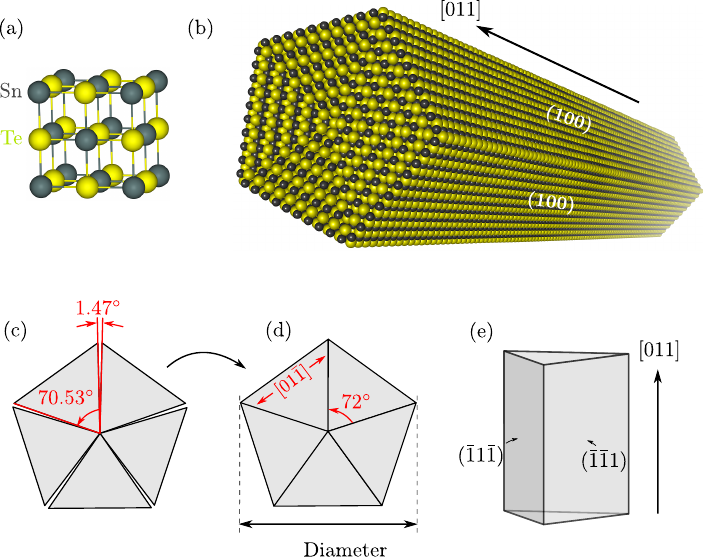}
		\subfloat{\label{fig:str_bulk}}
		\subfloat{\label{fig:str_NW}}
		\subfloat{\label{fig:cs_70}}
		\subfloat{\label{fig:cs_72}}
		\subfloat{\label{fig:triangle}}
		\caption{\footnotesize Geometry of the pentagonal IV--VI NW. (a) Rocksalt crystal structure. (b) Fivefold-twinned NW grown along the $[011]$ direction, composed of five wedge-shaped rocksalt domains separated by radial $\{111\}$ TPs and terminated by five $\{100\}$ outer facets. The TPs are either all cationic or all anionic and meet along the NW axis. (c) Cross section formed from undistorted domain wedges, for which the crystallographic angle between neighboring $\{111\}$ planes is $70.53^\circ$, leaving an angular deficit of $1.47^\circ$ per domain. (d) Regular pentagonal cross section obtained by deforming each wedge to an opening angle of $72^\circ$. (e) Side view of one trigonal-prismatic domain. }
		\label{fig:nwstr}
	\end{figure}

	SnTe-class IV--VI compounds crystallize in the rocksalt structure. We consider pentagonal NWs grown along the $[011]$ crystallographic direction and composed of five trigonal-prismatic rocksalt domains, as illustrated in Fig.~\ref{fig:nwstr}. Each such domain wedge is bounded externally by a $\{100\}$ facet and laterally by two $\{111\}$ TPs, while adjacent wedges are related by successive rotations of $72^\circ$ about the NW axis. The resulting ideal structure has point group $D_{5h}$. Owing to the sublattice arrangement of the rocksalt lattice, the five TPs necessarily have the same sublattice composition and are therefore either all cationic or all anionic. In the terminology of Ref.~\cite{deWit1972}, their common termination at the NW axis forms a star disclination.
	
	A scanning transmission electron microscopy image of a pentagonal $\mathrm{Pb}_{x}\mathrm{Sn}_{1-x}\mathrm{Te}$ NW viewed along the growth direction indicates an approximately $C_5$-symmetric structure with atoms present in the core region~\cite{hussain2024pentagonal}. Accordingly, in the main text we adopt an idealized occupied-core structure in which the five TPs meet at a single uncompensated atomic column along the NW axis. For cationic TPs this column consists of excess cations without corresponding anions, whereas for anionic TPs the sublattice composition is reversed, so that the structure is nonstoichiometric in both cases. Ref.~\cite{hussain2024pentagonal} also considered a stoichiometric core column obtained by introducing compensating atoms at interstitial positions near the axis. This high-symmetry configuration is unstable against unconstrained structural relaxation and evolves into a distorted core. We therefore do not use this core configuration in the main-text calculations. Its structural and electronic properties are examined in Appendix~\ref{app:str_stability}, which also considers a stoichiometric hollow-core structure obtained by removing the central column.
	
	The construction of a regular pentagonal cross section requires a small deformation of the rocksalt domains. The crystallographic angle between the two $\{111\}$ planes bounding an undistorted wedge is $\arccos(1/3)\approx70.53^\circ$, which is approximately $1.47^\circ$ smaller than the $72^\circ$ angle required to close the five-domain structure, corresponding to a total angular deficit of approximately $7.35^\circ$~\cite{deWit1972}. In the closed pentagonal geometry, this mismatch is accommodated by a uniaxial deformation of approximately $2\%$ along the local $[01\bar{1}]$ direction of each domain, as shown schematically in Figs.~\ref{fig:nwstr}(c) and \ref{fig:nwstr}(d). The associated transverse strain components expected from the elastic response can be estimated using the stiffness constants of SnTe and PbTe~\cite{Miller_1981,Zasavitskii_2004}. Together with the corresponding deformation potentials~\cite{Rabii_1969,Zasavitskii_2004}, these estimates give valley-dependent changes of the bulk gap below approximately $10\%$ for SnTe and $20\%$ for $\mathrm{Pb}_{0.4}\mathrm{Sn}_{0.6}\mathrm{Te}$. Since these corrections do not alter the band ordering relevant to the present analysis, we do not include a separate deformation-potential contribution in the subsequent TB calculations.
	
\begin{figure}[!htbp]
		\centering
		\includegraphics[width=0.47\textwidth]{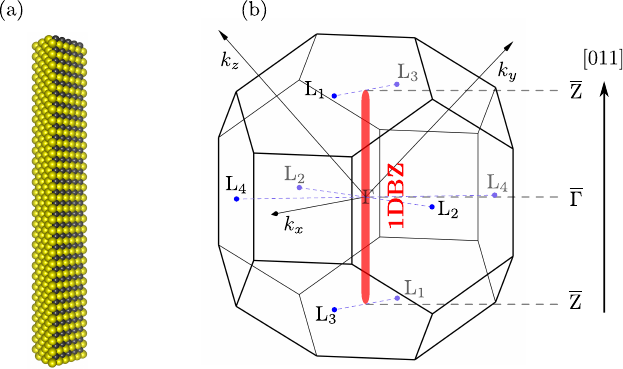}
		\caption{\footnotesize Reciprocal-space geometry of the pentagonal NW. (a) Perspective view of one trigonal-prismatic domain along the NW growth direction. (b) Three-dimensional fcc Brillouin zone and the one-dimensional Brillouin zone associated with translation along $[011]$. For each domain, the four bulk $L$ points project pairwise onto $\Gammabar$ at the center and $\Zbar$ at the boundary of the one-dimensional Brillouin zone. The five rotated domains share the same axial Brillouin zone.}
		\label{fig:bz}
	\end{figure}

	The NW is periodic along the $[011]$ growth direction. We keep its translation period at the unstrained bulk value $a_0\sqrt{2}/2$, where $a_0$ is the rocksalt lattice constant, thereby neglecting the axial strain induced by the transverse deformation through the Poisson effect. The unit cell contains two atomic layers. The corresponding one-dimensional Brillouin zone is shown in Fig.~\ref{fig:bz}. For each of the five trigonal-prismatic domains, the four bulk $L$ points project pairwise onto the two time-reversal-invariant momenta $\Gammabar$ and $\Zbar$. The low-energy NW spectrum is therefore concentrated near these two points.
	
	\section{Methods}
	\label{sec:methods}
	
	We calculate the electronic spectra of the pentagonal NWs using two TB models. Most calculations employ a simplified $p^3$ model for SnTe with the parameter set used in Ref.~\cite{Samadi2022_TP}, including nearest- and next-nearest-neighbor hopping. This model captures the essential low-energy band topology while allowing calculations for sufficiently large NW cross sections.
	
	For comparison with the experimentally relevant alloy, we also use a nearest-neighbor $sp^3d^5$ TB model~\cite{lent1986relativistic}. The parameters for $\mathrm{Pb}_{0.4}\mathrm{Sn}_{0.6}\mathrm{Te}$ are obtained within the virtual-crystal approximation from the corresponding PbTe and SnTe parameter sets.
	
	In both models, the Hamiltonian is evaluated using the atomic coordinates of the pentagonal geometry with the imposed deformation along the local $[01\bar{1}]$ direction. The resulting changes in bond orientation are included in the Slater--Koster matrix elements, while no separate deformation-potential correction or bond-length rescaling of the hopping parameters is applied, i.e., all parameters are retained at their bulk values. Transverse and axial strains arising from the elastic Poisson response are neglected. 
	
	Complementary DFT calculations are used to compare the stability and relaxation of alternative TP and core configurations, with the computational details given in Appendix~\ref{app:str_stability}.

	\section{Results}
	\label{sec:results}
	
	\subsection{Dirac core and surface modes in pentagonal nanowires}
	\label{sec:nwbs}
	
	\begin{figure}
		\centering
		\includegraphics[scale=1]{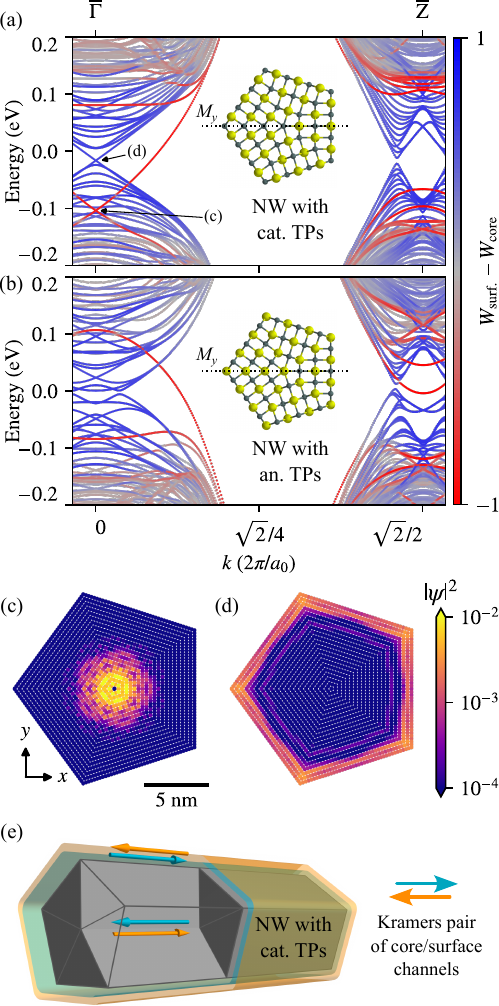}
		\caption{Band structures and wave-function localization in pentagonal SnTe NWs. Panels (a) and (b) show the spectra of 14~nm-thick (20-ring) NWs with cationic and anionic TPs, respectively; the inset schematics illustrate the corresponding three-ring structures for clarity. Red and blue color intensities indicate wave-function weight near the core and the outer surface. In panel (b), a local onsite-energy shift of 40~meV is applied to orbitals within 1~nm of the NW axis to move a topologically trivial core-localized subband away from the small surface gap; the unshifted spectrum is not shown. Panels (c) and (d) show the probability densities of the core- and surface-localized states forming the Dirac crossings in panel (a), while panel (e) schematically illustrates the resulting spatially separated channels in the NW with cationic TPs.}
		\label{fig:nw_slab_spectra}
	\end{figure}
	
	We consider SnTe NWs with either five cationic or five anionic TPs, and in both cases we use the idealized occupied-core structure introduced in Sec.~\ref{sec:IV-VI}. The NW size is specified by the number of pentagonal-prism rings, where one ring denotes a shell of one atomic layer (the central axial atoms are not included in the count). The structural schematics in Fig.~\ref{fig:nw_slab_spectra} show three-ring NWs for clarity, whereas the calculated spectra correspond to 20-ring NWs, with a thickness of approximately 14~nm measured along the pentagon diagonal.
	
	The spectra calculated within the bulk-gap energy window are shown in Fig.~\ref{fig:nw_slab_spectra}(a) for cationic TPs and in Fig.~\ref{fig:nw_slab_spectra}(b) for anionic TPs. The states are colored according to the difference between $W_{\mathrm{surf.}}$, defined as the probability density within the two outermost atomic layers, and $W_{\mathrm{core}}$, defined as the probability density within a distance $3.7a_0$ of the NW axis. Since the bulk gap of the simplified SnTe parametrization is approximately 330~meV, most of the states in the presented energy window originate from the TCI surface spectrum. In a NW geometry these states wrap around the surface and are quantized into multiple one-dimensional subbands. Throughout, we use the term surface to refer collectively to all five faces and all five hinges of the pentagonal NW.
	
	For cationic TPs, the spectrum contains two linearly dispersing crossings at $k=0$, i.e., at the $\Gammabar$ point. One crossing is formed by states localized near the NW axis, as demonstrated by the probability density in Fig.~\ref{fig:nw_slab_spectra}(c), while the second is formed by states distributed over the outer surface, with enhanced weight at the hinges as shown in Fig.~\ref{fig:nw_slab_spectra}(d). The two crossings have analogous low-energy structures but occur on spatially separated boundaries of the wire. We have verified numerically that the effects of finite-size hybridization between the core- and surface-localized states become negligible once the wire thickness exceeds their localization lengths. In the simplified model, a thickness of approximately 14~nm is sufficient for the two crossings to be effectively gapless. Although this thickness is smaller than the 40--150~nm range reported experimentally for $\mathrm{Pb}_{1-x}\mathrm{Sn}_{x}\mathrm{Te}$ NWs~\cite{hussain2024pentagonal}, it already spatially separates the core and surface states in the simplified SnTe model.
	
	The NW with anionic TPs behaves qualitatively differently. Neither the core nor the surface sector develops a massless Dirac crossing near $\Gammabar$, even when the radius is increased beyond the 20-ring value used for the calculation shown in Fig.~\ref{fig:nw_slab_spectra}(b). The corresponding subbands may nevertheless overlap in energy, so the full spectrum need not remain globally gapped near $\Gammabar$.

	Near $\Zbar$, the core and surface spectra are gapped for both TP types. We have verified numerically that these gaps decrease overall with increasing perimeter but do not evolve into protected Dirac crossings when the core and surface states become spatially decoupled. Since the sector near $\Zbar$ does not contribute to the one-dimensional gapless modes that are the focus of the present work, its detailed analysis in terms of closed-surface momentum quantization, mirror-related helical modes, and a low-energy theory is deferred to Appendix~\ref{app:zbar}. The remainder of the main text therefore concentrates on the $\Gammabar$ sector.
	
	\subsection{Origin and protection of the $\Gammabar$ modes}
	\label{sec:gamma_protection}
	\label{sec:kp_models}
	\label{sec:cyl_slab_TP}
	
	\begin{figure}
		\centering
		\includegraphics[scale=1]{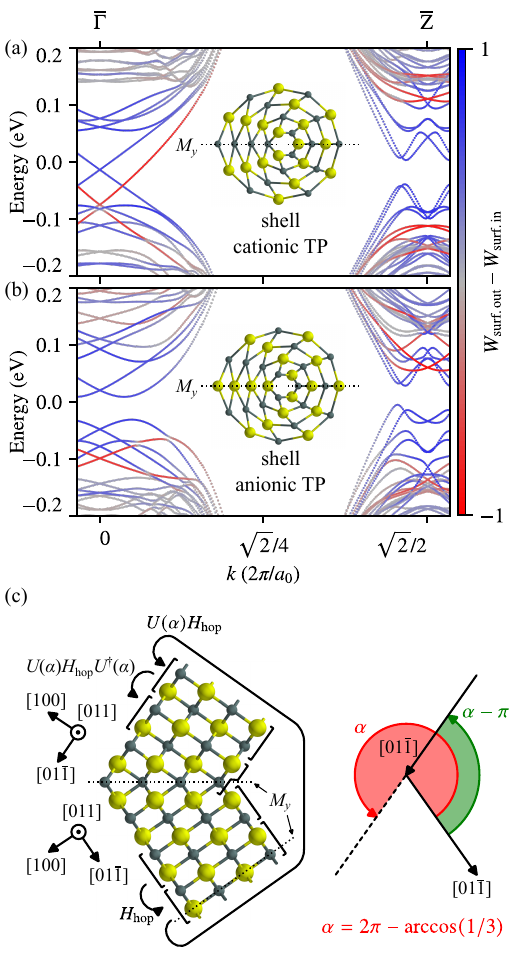}
		\caption{Band structures of cylindrical SnTe shells containing a single (a) cationic and (b) anionic TP. The calculated shells are 18 atomic layers thick and contain 40 atoms around the outer circumference. Red and blue color intensities indicate wave-function weight near the inner and outer surfaces, respectively. The inset schematics show four-layer shells with 12 atoms around the outer circumference; the twisted periodic boundary is represented by bending each shell into its closed form to make the common $M_y$ mirror plane explicit. Panel (c) illustrates the construction of the twisted boundary condition using a four-layer shell with 14 atoms around the outer circumference.}
		\label{fig:TP_slab_spectra}
	\end{figure}
	
	The spatial localization of the two crossings in the NW with cationic TPs suggests a boundary-mode interpretation. The core crossing is associated with a mode bound to the one-dimensional defect where the five TPs meet, whereas its partner is localized at the outer surface. The core and outer surface can thus be viewed as the two boundaries of an effective two-dimensional time-reversal-invariant subsystem, analogous to the opposite edges of a quantum spin Hall ribbon. The resulting spatially separated channels are illustrated in Fig.~\ref{fig:nw_slab_spectra}(e).
	
	To identify the microscopic origin of these modes, we first consider a minimal atomistic system that retains the distinction between the two TP types. It is obtained by introducing a single $(111)$ TP into a cylindrical SnTe shell. The shell is constructed by joining two defect-free $(100)$ slabs along the TP and connecting their remaining free ends through a regular $(01\bar{1})$ atomic plane, as shown in Fig.~\ref{fig:TP_slab_spectra}(c). This geometry may be viewed as a hollow-core partial wedge disclination in the sense of Ref.~\cite{deWit1972}, with net lattice rotation
	\begin{equation}
		\alpha=2\pi-\arccos(1/3).
		\label{eq:single_tp_alpha}
	\end{equation}
	Following the disclination construction of Ref.~\cite{Geier2021}, we connect the two free boundaries through the rotation-twisted hopping
	\begin{equation}
		U(\alpha) H_{\mathrm{hop}},
		\label{eq:single_tp_hopping}
	\end{equation}
	where $H_{\mathrm{hop}}$ is the hopping matrix between equivalent regular atomic planes in the surrounding slab and $U(\alpha)$ rotates the full spin--orbital basis counterclockwise by the angle $\alpha$ about the $[011]$ axis. This choice makes the stitching locally homogeneous with the slab away from the TP, so that the seam introduces only the accumulated lattice rotation rather than an additional microscopic interface. The hopping amplitudes are otherwise taken from the corresponding unbent geometry, so that the construction isolates the effects of the TP and the boundary condition from the large strain that would accompany a literal bending of the crystal.
	
	For the calculations in Fig.~\ref{fig:TP_slab_spectra}, the shell contains 18 atomic layers, two fewer than the 20 rings used for the corresponding pentagonal NW. The two innermost layers are omitted as a technical choice to avoid ambiguity in defining the boundary condition at the inner surface, while the outer perimeter is chosen to equal one fifth of the full NW perimeter. A shell containing a cationic TP exhibits two Dirac crossings near $\Gammabar$, one localized at the outer surface and the other at the inner surface, as shown in Fig.~\ref{fig:TP_slab_spectra}(a). By contrast, the shell containing an anionic TP exhibits no corresponding crossings, as shown in Fig.~\ref{fig:TP_slab_spectra}(b). These results show directly that the Dirac crossings do not require the complete fivefold geometry but already emerge in a system containing a single cationic TP. The crossings can therefore be interpreted as modes associated with individual TPs, whose terminations lie on the outer and inner boundaries of the shell. In the pentagonal NW, the five outer terminations coincide with the hinges between neighboring facets, while the corresponding inner terminations collapse onto the common NW axis.
	
	Moreover, the connection to the single-TP shell can be made formal by decomposing the full pentagonal NW Hamiltonian into eigenspaces of the spinful rotation $C_5$, as derived microscopically in Appendix~\ref{app:c5_basis}. The allowed $C_5$ eigenvalues are
	\begin{equation}
		\lambda_\nu=e^{-i\pi(2\nu+5)/5},
		\qquad \nu=-2,-1,0,1,2,
		\label{eq:c5_eigenvalues}
	\end{equation}
	with $\lambda_0=-1$. After accounting for the small strain deformation required to close the pentagonal geometry, the Hamiltonian projected onto the $C_5=-1$ eigenspace reduces to the Hamiltonian of the cylindrical shell containing a single TP. The single-TP shell is therefore not merely a diagnostic geometry, but its spectrum is embedded in the full NW spectrum as the $C_5=-1$ sector. In particular, the two Dirac crossings of the cationic shell correspond to the core and surface crossings of this rotational block, further supporting their association with the individual TPs.
	
	The single-TP dependence on the TP sublattice type is consistent with the earlier analysis of isolated TPs and twinning superlattices in Ref.~\cite{Samadi2022_TP}, which found protected side-surface crossings near the $\Gammabar$ analogue for cationic TPs but not for anionic TPs. That work characterized the twin-plane systems by mirror Chern numbers and also found additional mirror-protected crossings near the $\Mbar$ points for both TP types. In the NW geometry, the $\Mbar$-analogue states occur near $\Zbar$. Their modified behavior upon closing the surface around the NW perimeter is discussed in Appendix~\ref{app:zbar}. The mirror Chern numbers of Ref.~\cite{Samadi2022_TP} characterize the combined mirror-resolved surface spectrum, including crossings in both momentum regions, and should therefore not be interpreted as direct indices of the isolated $\Gammabar$ sector considered here.
	
	\begin{figure}
		\centering
		\includegraphics[scale=1]{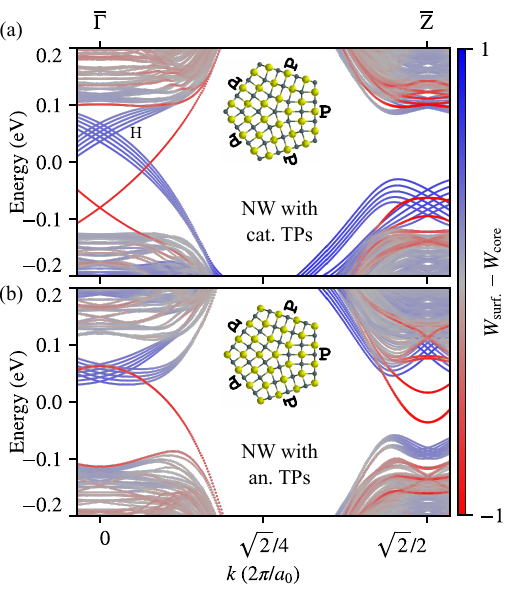}
		\caption{Band structures of pentagonal SnTe NWs after applying time-reversal-preserving perturbations to the outermost atomic layers. Panels (a) and (b) correspond to NWs with cationic and anionic TPs, respectively. The inset schematics indicate the perturbed surface regions, with $P$ denoting perturbations that break the vertical $M_y$ mirror symmetries of the NW in the surface region.}
		\label{fig:pert_NW_spectra}
	\end{figure}
	
	The single-TP shell identifies a helical pair associated with the outer edge of each cationic TP. In the full pentagonal NW, the corresponding five TP-edge pairs can be isolated from the remaining TCI-derived surface subbands by introducing a perturbation that gaps the extended surface spectrum. We introduce a time-reversal-preserving perturbation confined to the outermost atomic layer of the five NW faces. The perturbation consists of onsite terms that are homogeneous within each face and is chosen to break the horizontal mirror symmetry $M_z$, perpendicular to the NW axis, together with the five vertical mirror symmetries containing the TPs, represented by $M_y$ and its $C_5$-related counterparts. Weak electrostatic potentials of different magnitudes are then applied at the five TP edges, breaking the $C_5$ rotational symmetry and separating the resulting hinge channels in energy. The NW thickness is the same as in Sec.~\ref{sec:nwbs}.
	
	The resulting spectra are shown in Figs.~\ref{fig:pert_NW_spectra}(a,b). In the NW with cationic TPs, five helical Kramers pairs emerge near $\Gammabar$, each localized at one TP edge, whereas no corresponding modes appear in the anionic wire. Restoring the unperturbed surface spectrum allows these five outer-boundary channels to hybridize through the conducting faces. Because their number is odd, four combinations can acquire masses while one helical pair remains gapless. The same odd-channel mechanism is expected to apply at the NW core, where the five TP-associated channels meet and can hybridize directly, leaving one helical Kramers pair.
	
	To make this argument explicit, we construct the low-energy theory of five helical Kramers pairs in the idealized $D_{5h}$-symmetric wire. We choose a basis that diagonalizes $C_5$ and the mirror reflection $M_z$ perpendicular to the NW axis. In the $\lambda_0=-1$ sector the symmetry operators may be represented as
	\begin{equation}
		\begin{split}
			C_5^{\{0\}}&=-\mathbb{I}_{2\times2},
			\qquad M_y^{\{0\}}=-is_y,\\
			M_z^{\{0\}}&=-is_z,
			\qquad \Theta^{\{0\}}=is_yK,
		\end{split}
		\label{eq:gamma_symmetry_minus_one}
	\end{equation}
	where $s_{x,y,z}$ are Pauli matrices in the two-component helical subspace and $K$ denotes complex conjugation. The remaining sectors occur in complex-conjugate pairs $(\lambda_\nu,\lambda_{-\nu})$, with $\nu=1,2$, and the corresponding symmetry matrices are
	\begin{equation}
		\begin{split}
			C_5^{\{\nu,-\nu\}}&=
			\begin{pmatrix}
				\lambda_\nu \mathbb{I}_{2\times2} &0\\
				0&\lambda_{-\nu} \mathbb{I}_{2\times2}
			\end{pmatrix},\\
			M_y^{\{\nu,-\nu\}}&=
			\begin{pmatrix}
				0&-is_y\\
				-is_y&0
			\end{pmatrix},\\
			M_z^{\{\nu,-\nu\}}&=
			\begin{pmatrix}
				-is_z&0\\
				0&-is_z
			\end{pmatrix},\\
			\Theta^{\{\nu,-\nu\}}&=
			\begin{pmatrix}
				0&is_y\\
				is_y&0
			\end{pmatrix}K.
		\end{split}
		\label{eq:gamma_symmetry_paired}
	\end{equation}
	The construction of the $C_5$ eigenbasis and the projected symmetry representations used above are given in Appendix~\ref{app:c5_basis}.
	
\begin{figure*}
		\centering
		\includegraphics[scale=1]{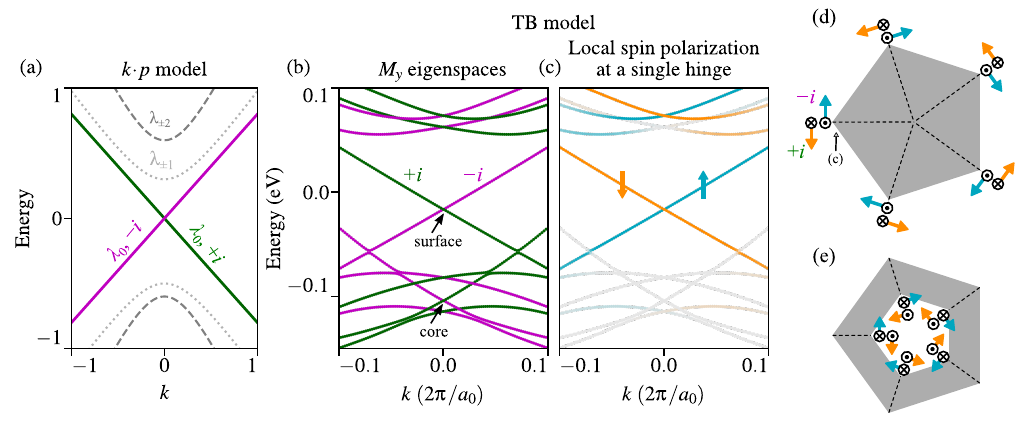}
		\caption{Low-energy spectrum and spin structure of a NW with cationic TPs. (a) Representative spectrum of the symmetry-based low-energy Hamiltonian~\eqref{eq:ham_c5} near $\Gammabar$. The solid green and magenta bands form the gapless $C_5=-1$ sector and belong to the $M_y=+i$ and $M_y=-i$ eigenspaces, respectively. The dashed bands form the two massive complex-conjugate pairs of rotational sectors. (b,c) Close-ups of the TB spectrum near $\Gammabar$ showing all $C_5$ sectors, resolved according to (b) the $M_y$ eigenspaces and (c) the local spin polarization evaluated on a representative TP-edge atomic column, marked by the black arrow in panel (d). Green and magenta in panel (b) denote the $M_y=+i$ and $M_y=-i$ eigenspaces, respectively. Orange and cyan in panel (c) indicate opposite local spin polarizations. (d,e) Schematic spin textures of the five TP-edge components at the outer boundary and at an inner boundary representing the core mode, respectively. The colored arrows indicate the local spin polarizations perpendicular to the corresponding TP planes, while the circle-dot and circle-cross symbols denote opposite propagation directions along the NW axis. The reversed spin--momentum locking between the two boundaries reflects the opposite helicities of the surface and core modes. In each case, the five local polarization vectors are related by $C_5$ and sum to zero.}
		\label{fig:kp_and_spin}
	\end{figure*}

	Keeping terms through first order in the momentum $k_z$ measured from $\Gammabar$, the most general Hamiltonian in a rotational sector is
	\begin{equation}
		H^{\{\nu\}}(k_z)=\epsilon_\nu s_0+\Delta_\nu s_z+v_\nu k_zs_y,
		\label{eq:ham_c5}
	\end{equation}
	with the constraints
	\begin{equation}
		\epsilon_\nu=\epsilon_{-\nu},
		\qquad v_\nu=v_{-\nu},
		\qquad \Delta_\nu=-\Delta_{-\nu}.
		\label{eq:gamma_parameter_constraints}
	\end{equation}
	The self-conjugate sector $\nu=0$ therefore satisfies $\Delta_0=0$, and its spectrum necessarily contains a linearly dispersing crossing. Within this $C_5=-1$ sector, the two counterpropagating branches can be chosen as eigenstates of $M_y$ with eigenvalues $+i$ and $-i$. They are shown in green and magenta, respectively, in Fig.~\ref{fig:kp_and_spin}(a). By contrast, the $\nu=\pm1$ and $\nu=\pm2$ sectors are generally massive, with the spectra of each conjugate pair remaining degenerate. The five original helical pairs thus reorganize into two gapped conjugate pairs of rotational sectors and one ungapped $C_5=-1$ sector.
	
	The two-branch low-energy spectrum in Fig.~\ref{fig:kp_and_spin}(a) describes either the core or the surface crossing separately. In the full TB spectrum shown in Fig.~\ref{fig:kp_and_spin}(b), each of the two previously identified crossings consists of counterpropagating branches belonging to the opposite $M_y=+i$ and $M_y=-i$ eigenspaces, in agreement with the low-energy theory.
	
	The mirror eigenvalue also constrains the local spin expectation value of the outer-surface mode. At the representative TP-edge atomic column marked in Fig.~\ref{fig:kp_and_spin}(d), $M_y$ symmetry restricts the spin polarization to the direction perpendicular to the corresponding TP plane. Figure~\ref{fig:kp_and_spin}(c) shows that this local spin component has opposite signs on the two counterpropagating branches. These branches are eigenstates of $M_y$, rather than spin eigenstates, and the magnitude of their local spin polarization is therefore not quantized. The five TP-edge components are related by successive $C_5$ rotations and have the polarization pattern shown schematically in Fig.~\ref{fig:kp_and_spin}(d). Their vector sum vanishes, so the outer-surface Dirac mode carries no net spin polarization in the ideal $D_{5h}$-symmetric NW despite being locally spin polarized at each TP edge. A single strongly spin-polarized hinge channel is recovered when the coupling between the five TP-edge components is suppressed, as in the perturbation calculation of Fig.~\ref{fig:pert_NW_spectra}(a).
	
	Finally, we note that, since the ungapped $C_5=-1$ sector forms a single helical Kramers pair, a time-reversal-preserving perturbation may shift or deform its dispersion but cannot generate a mass without coupling it to another helical pair. Exact fivefold and mirror symmetries are therefore useful for identifying the protected sector but are not required for the persistence of the Dirac modes. The $\Gammabar$ crossing remains stable under nonmagnetic crystalline distortions as long as the bulk gap and the gaps of the two-dimensional states localized on the extended TPs remain open, and the core and outer-surface channels are not strongly mixed.
	
	\subsection{Origin and robustness of the core mode}
	\label{sec:core_robustness}
	
	While the symmetry-based low-energy model predicts the core crossing, its microscopic origin can also be conceptualized by considering the hollow NW shown in Fig.~\ref{fig:kp_and_spin}(e). In this geometry, the five TPs terminate at an inner surface as well as at the outer one. For cationic TPs, the inner boundary supports the same five helical pairs near $\Gammabar$ as the outer boundary, but with opposite helicity because the surface normal is reversed. Their hybridization produces a single gapless $C_5=-1$ pair on the inner surface, in direct analogy with the outer-surface mode.
	
	The inner perimeter may be reduced continuously until it becomes the microscopic core region of the filled wire. This deformation preserves time-reversal symmetry, the relevant rotational sector, and the inverted band gap of the surrounding material. It therefore cannot remove the isolated helical pair without closing a gap or bringing it into contact with another pair. Shrinking the perimeter increases the energy spacing between the remaining core-localized subbands, leaving the core crossing in Fig.~\ref{fig:nw_slab_spectra}(a) more spectrally isolated than the corresponding outer-surface crossing.
	
	The contraction construction does not by itself exclude additional core-localized bands associated with an occupied central atomic column. We therefore analyze the contribution of the central column separately using rotational symmetry in Appendix~\ref{app:c5_case}. An atom located exactly on the NW axis is fixed in real space by $C_5$, so the rotational eigenvalue of an $s$ or $p$ orbital is determined entirely by its spin and orbital angular momentum. None of the spinful $s$ or $p$ states on the central atom carries the eigenvalue $\lambda_0=-1$ of the Dirac crossing. Consequently, in the simplified $p$-orbital model, the core mode has exactly zero weight on the central column. In the realistic $sp^3d^5$ model, a symmetry-allowed central-atom contribution can arise only through the $d$ orbitals and is expected to remain weak near the band gap. The crossing is therefore formed primarily from the rings of atoms surrounding the NW axis and persists in both occupied- and hollow-core calculations.
	
	The selection rule associated withe core column further distinguishes the helical core modes investigated here from the core-chain band reported in Ref.~\cite{hussain2024pentagonal}. The latter is localized predominantly on the central atomic column and forms a twofold-degenerate subband. As shown in Appendix~\ref{app:str_stability}, our DFT calculations recover this band for the constrained high-symmetry stoichiometric core configuration considered in Ref.~\cite{hussain2024pentagonal}, but it disappears upon unconstrained structural relaxation. The core-chain band is therefore microscopically distinct from the boundary-derived core mode discussed here.
	
	Lastly, we comment on the spin structure of the helical core mode. An argument analogous to that for the outer-surface mode applies. Viewing the core state as the continuation of the inner-boundary mode as the hollow core is contracted, its five symmetry-related components inherit the same mirror-constrained local spin polarization as the outer-surface components, but with reversed helicity. Their polarization vectors are related by successive $C_5$ rotations and sum to zero, so the core Dirac mode likewise carries no net spin polarization in the ideal $D_{5h}$-symmetric NW.

	\subsection{Realistic $\mathrm{Pb}_{0.4}\mathrm{Sn}_{0.6}\mathrm{Te}$ nanowires}
	\label{sec:realistic_model}
	
	\begin{figure*}[!htbp]
		\centering
		\includegraphics[scale=1]{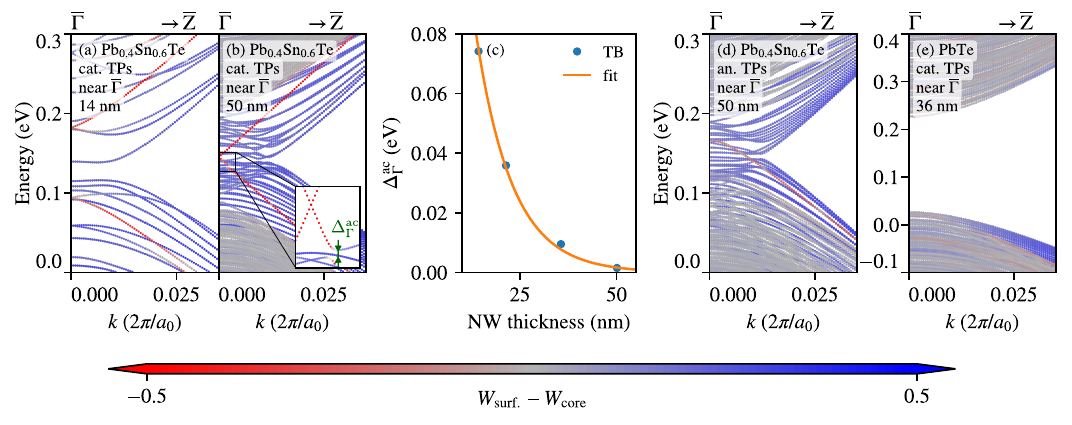}
		\caption{Band structures and finite-size scaling obtained with the $sp^3d^5$ model. (a,b) Spectra of $\mathrm{Pb}_{0.4}\mathrm{Sn}_{0.6}\mathrm{Te}$ NWs with cationic TPs at thicknesses of 14 and 50~nm, respectively. The inset in panel (b) magnifies the vicinity of the core and surface Dirac points and the core--surface anticrossing near $\Gammabar$. (c) Hybridization-induced anticrossing $\Delta_{\Gamma}^{\mathrm{ac}}$ as a function of NW thickness. Symbols show the TB results and the line is the exponential fit of Eq.~\eqref{eq:realistic_gap_scaling}. (d) Spectrum of a 50-nm-thick $\mathrm{Pb}_{0.4}\mathrm{Sn}_{0.6}\mathrm{Te}$ NW with anionic TPs. (e) Spectrum of a 36-nm-thick topologically trivial PbTe pentagonal NW with cationic TPs. Red and blue color intensities indicate wave-function weight near the core and the outer surface, respectively.}
		\label{fig:lent_bs}
	\end{figure*}
	
	The experimentally fabricated pentagonal NWs are ternary alloys with an inverted gap smaller than that of SnTe. We therefore perform nearest-neighbor $sp^3d^5$ calculations for $\mathrm{Pb}_{0.4}\mathrm{Sn}_{0.6}\mathrm{Te}$, whose composition is close to that reported in Ref.~\cite{hussain2024pentagonal}. The calculation retains the approximate treatment of strain adopted in the simplified model. Within the virtual-crystal parametrization, the alloy lies in the TCI regime with an inverted bulk gap of approximately 104~meV. The smaller gap increases the localization lengths of the boundary states and makes finite-size hybridization more pronounced.
	
	Figure~\ref{fig:lent_bs} summarizes the resulting spectra near $\Gammabar$. In the 14-nm-thick cationic-TP NW shown in panel (a), the core and outer-surface states overlap strongly and the two nominally helical modes form a visible avoided crossing. At a thickness of 50~nm, shown in panel (b), their spatial overlap is strongly suppressed: the core-localized bands approach a massless linear dispersion, while the surface-localized partner crossing becomes visible among the more densely spaced TCI-derived surface subbands. The inset resolves the small residual anticrossing near the core- and surface-localized Dirac points. This thickness lies within the experimentally reported range.
	
	To quantify the decoupling of the two cationic-TP modes, we extract their hybridization-induced anticrossing from calculations at thicknesses of 14, 21, 36, and 50~nm, corresponding to 20, 30, 50, and 70 rings. The results are shown in Fig.~\ref{fig:lent_bs}(c) and follow the exponential dependence
	\begin{equation}
		\Delta_{\Gamma}^{\mathrm{ac}}\propto e^{-\gamma d},
		\qquad \gamma=0.1065~\mathrm{nm}^{-1},
		\label{eq:realistic_gap_scaling}
	\end{equation}
	where $d$ is the NW thickness. The exponential decay reflects the decreasing overlap between states localized at the core and the outer surface. The quantity $\Delta_{\Gamma}^{\mathrm{ac}}$ is not, in general, a global gap in the full NW spectrum, since topologically trivial surface subbands may overlap the same energy range both near $\Gammabar$ and $\Zbar$. It characterizes only the anticrossing produced by hybridization of the core and surface modes and is distinct from the confinement-controlled energy spacings of the other surface subbands.
	
	Panel (d) confirms within the $sp^3d^5$ model the distinction between cationic and anionic TPs already found using the simplified model. For anionic TPs, calculations over the same sequence of NW thicknesses show that increasing the thickness spatially separates the core and surface states but does not produce a Dirac crossing. The core dispersion remains topologically trivial. The representative 50~nm spectrum contains several overlapping subbands and is therefore not globally gapped, but no helical Dirac crossing develops.
	
	The material-specific calculations also preserve the qualitative distinction between the $\Gammabar$ and $\Zbar$ momentum sectors. No protected crossing appears near $\Zbar$ for either TP type, even in 50~nm NWs for which the $\Gammabar$ core and surface modes are nearly decoupled. The corresponding spectra are shown in Appendix~\ref{app:zbar}.
	
	Finally, Fig.~\ref{fig:lent_bs}(e) isolates the role of bulk band inversion by showing a topologically trivial PbTe NW with the pentagonal cationic-TP geometry. The core and surface Dirac dispersions are absent in this case. Together with the anionic-TP results, this control calculation shows that the core and surface Dirac crossings emerge only when the bulk band structure is inverted and the TPs are cationic. Altogether, the $sp^3d^5$ model confirms that these helical modes are not artifacts of the simplified Hamiltonian and become well developed at experimentally accessible NW thicknesses.
	
	\section{Discussion and conclusions}
	\label{sec:conclusion}
	
	We have shown that pentagonal NWs made of SnTe-class compounds can host spatially separated helical Dirac modes near the NW core and at the outer surface. These modes emerge when the bulk band structure is inverted and the five radial TPs are cationic, whereas the corresponding spectra of NWs with anionic TPs remain topologically trivial. The core and surface crossings can be understood as spatially separated helical boundary modes of an effective two-dimensional $\mathbb{Z}_2$ topological subsystem. Microscopically, the spectrum results from the coupling of the boundary modes associated with the five twin-related sectors, whose hybridization leaves a single time-reversal-protected Dirac crossing at the core and a corresponding crossing at the outer surface. The cationic $\Gammabar$ modes are closely related to the side-surface states predicted previously for TP slabs and twinning superlattices~\cite{Samadi2022_TP}, whereas the different behavior of the corresponding $\Zbar$ sector reflects the closure of the surface around the NW perimeter.
	
	The Dirac crossings are expected to be robust against non-magnetic disorder and moderate structural distortions, provided that these perturbations do not close the protecting gap or strongly couple the core and outer surface. In particular, the core mode persists in the different idealized microscopic realizations of the NW axis considered here, including structures with an occupied or hollow core. Its existence is consequently not tied to dangling bonds of the central atomic column or to a particular microscopic termination of the five TPs, but follows from the topology of the surrounding nanowire structure.
	
	In the realistic $\mathrm{Pb}_{0.4}\mathrm{Sn}_{0.6}\mathrm{Te}$ model, well-developed core and surface Dirac modes emerge at NW thicknesses larger than approximately $50$~nm, within the range of experimentally fabricated pentagonal NWs~\cite{hussain2024pentagonal}. The core mode is strongly confined around the NW axis and is also spectrally separated from the remaining core-localized subbands. Its surface counterpart, by contrast, coexists and hybridizes with the additional states inherited from the TCI surface spectrum. The core crossing therefore provides an experimentally viable realization of a separated helical Dirac mode propagating along the one-dimensional defect.
	
	Similar one-dimensional topological modes have been predicted in a variety of material systems at crystalline defects including dislocations~\cite{ran2009,Slager2014}, partial dislocations and stacking-fault terminations~\cite{Queiroz2019,Tuegel2019,Naselli2022}, and wedge-disclination axes~\cite{Geier2021}. Dislocation-bound topological modes have also been observed experimentally~\cite{nayak2019resolving}. Fabricating isolated defects with the required structure is challenging, particularly when the defect must extend coherently through the length of a nanostructure. In the pentagonal geometry considered here, the relevant defect is instead an intrinsic element of the NW morphology, produced naturally by the five radially arranged TPs meeting along the growth axis. Second-order topological insulators provide another realization of helical modes in NWs, where the bulk and surfaces can be gapped while helical Kramers pairs remain at selected hinges~\cite{langbehn2017reflection,schindler2018higher}. A distinct route is available in strong-topological-insulator NWs, where an axial half-flux quantum produces a gapless one-dimensional helical surface mode~\cite{zhang2010anomalous,hong2014one,cho2015aharonov,jauregui2016magnetic}. Unlike this flux-induced surface mode, the channels considered here arise intrinsically from the crystalline structure of the NW.

	The core and surface channels form the nanowire analogue of the two edges of a quantum spin Hall ribbon. This analogy suggests possible applications involving superconductivity, since proximity coupling of helical channels to an $s$-wave superconductor, together with a locally induced time-reversal-breaking gap, provides a standard route to topological superconducting interfaces and Majorana bound states~\cite{FuKane2008,FuKane2009,XuFu2010}. In the present geometry, the spatial separation between the core and surface channels may permit superconducting and magnetic perturbations to couple differently to the two sectors, although achieving the required selectivity and isolating a single low-energy channel on the NW surface would require further development.
	
	A central experimental challenge is the control of the TP sublattice type. The \textit{ab initio} calculations presented in Appendix~\ref{app:str_stability} indicate that, for the idealized SnTe structures considered there, the anionic variant is energetically favored. This preference need not be universal and may depend on growth conditions, stoichiometry, or alloy composition. Further work is therefore needed to determine whether cationic TPs can be stabilized during growth and to identify which SnTe-class compounds and solid solutions can form stable pentagonal NWs while retaining an inverted bulk band structure. It will also be important to establish the microscopic core structures realized in sufficiently large NWs and to characterize how disorder and electrostatic band bending affect the localization, spectral isolation, and accessibility of the core and surface modes.
	
	\begin{acknowledgments}
		We thank Carmine Autieri, Giuseppe Cuono, Piotr Dziawa, Ion Cosma Fulga, Ghulam Hussain, Alexander Lau, and Jakub Polaczyński for useful discussions.  
		
		We acknowledge financial support by the Polish National Science Centre (NCN) Grant under project No. 2016/23/B/ST3/03725. S.S. and R.B. also thank the financial support of the NCN grant under the IMPRESS-U project No. 2023/05/Y/ST3/00191. M.A.Ch. thanks The Polish National Center for Research and Development (grant no. EIG CONCERT-JAPAN/9/56/AtLv-AlGaN/2023). Computations were carried out using the computers of Centre of Informatics Tricity Academic Supercomputer \& Network.
		
	\end{acknowledgments}

	\appendix
	
	\section{Central-atom contribution to the core states}
	\label{app:c5_case}
	
	The core Dirac crossing of the NW with cationic TPs belongs to the self-conjugate rotational sector $\nu=0$, with $C_5$ eigenvalue $\lambda_0=-1$. The idealized occupied-core structure contains an atomic column exactly on the rotation axis, whose atoms are fivefold coordinated rather than sixfold coordinated as in the bulk rocksalt lattice. This reduced coordination could in principle generate dangling-bond-derived states near the core. Because the axial sites are fixed by $C_5$, their rotational eigenvalues are determined entirely by their spin and orbital angular momenta, providing a direct test of whether orbitals on the central column can contribute to the Dirac crossing.
	
	For a product state $|m_s,m_l\rangle$, where $m_s=\pm 1/2$ and $m_l$ denote, respectively, the spin and orbital angular-momentum projections along the NW axis,
	\begin{equation}
		\hat C_5|m_s,m_l\rangle
		=
		e^{-i(m_s+m_l)2\pi/5}|m_s,m_l\rangle.
	\end{equation}
	Membership in the $\nu=0$ sector therefore requires
	\begin{equation}
		m_s+m_l\equiv\frac{5}{2}\pmod{5}.
		\label{eq:c5-central-condition}
	\end{equation}
	For $m_s=1/2$ and $m_s=-1/2$, this condition requires $m_l=2$ and $m_l=-2$, respectively, modulo $5$. These values are unavailable to axial $s$ and $p$ orbitals, for which $|m_l|<2$. Thus, the $C_5=-1$ sector consequently has exactly zero weight on the axial $s$ and $p$ orbitals. Within the simplified $p^3$ TB model, the core Dirac state must therefore be formed from orbitals on the surrounding atomic rings rather than from the central atomic column.
	
	The selection rule does not exclude all higher-angular-momentum orbitals. In particular, the $m_l=\pm2$ components of $d$ orbitals can satisfy Eq.~\eqref{eq:c5-central-condition}, so an $sp^3d^5$ model may in principle contain a symmetry-allowed contribution from axial $d$ orbitals. The low-energy bands of SnTe-class compounds are, however, predominantly derived from $p$ orbitals~\cite{Littlewood2010}, and a large axial $d$-orbital contribution near the band gap is therefore not expected.

	\section{Nanowire Hamiltonian in the eigenbasis of the $C_5$ rotation}
	\label{app:c5_basis}
	
	This appendix first complements Appendix~\ref{app:c5_case} by showing how off-axis $s$ and $p$ orbitals arranged in five-site $C_5$ orbits can contribute to the $C_5=-1$ sector through their real-space angular character. We then develop the microscopic rotational decomposition of the pentagonal NW Hamiltonian, verify it in the calculated spectra, relate the $C_5=-1$ block to the cylindrical shell containing a single TP, and derive the projected symmetry representations used in the five-channel low-energy theory. Orbitals located exactly on the rotation axis are not included in the following constructions because their real-space positions are fixed under $C_5$. Their rotational selection rules are analyzed in Appendix~\ref{app:c5_case}.
	
	\subsection{Off-axis orbital content of the $C_5=-1$ sector}
	
	Every off-axis atomic site belongs to an orbit of five sites related by successive $C_5$ rotations. We label these lattice sites by $n=0,\ldots,4$ in counterclockwise order when the cross section is viewed from the positive end of the NW axis toward the origin, so that an active counterclockwise rotation maps $n$ onto $n+1$, with the index understood modulo five. The lattice-site part of the rotation is then represented by
	\begin{equation}
		P_5=
		\begin{pmatrix}
			0&0&0&0&1\\
			1&0&0&0&0\\
			0&1&0&0&0\\
			0&0&1&0&0\\
			0&0&0&1&0
		\end{pmatrix}.
		\label{eq:c5_offaxis_shift}
	\end{equation}
	Its normalized eigenvectors may be written as
	\begin{equation}
		|m_\mathrm{lat.}\rangle
		=\frac{1}{\sqrt{5}}
		\begin{pmatrix}
			1\\
			e^{im_\mathrm{lat.}2\pi/5}\\
			e^{2im_\mathrm{lat.}2\pi/5}\\
			e^{3im_\mathrm{lat.}2\pi/5}\\
			e^{4im_\mathrm{lat.}2\pi/5}
		\end{pmatrix},
		\qquad m_\mathrm{lat.}=0,\pm1,\pm2,
	\end{equation}
	with
	\begin{equation}
		P_5|m_\mathrm{lat.}\rangle=e^{-im_\mathrm{lat.}2\pi/5}|m_\mathrm{lat.}\rangle.
	\end{equation}
	
	Combining the lattice-site orbit with a spin state $|S_z=m_s\rangle$ and an internal orbital state $|m_l\rangle$ gives
	\begin{equation}
		|m_s,m_l,m_\mathrm{lat.}\rangle
		=|S_z=m_s\rangle\otimes|m_l\rangle\otimes|m_\mathrm{lat.}\rangle,
	\end{equation}
	which transforms according to
	\begin{equation}
		\hat C_5|m_s,m_l,m_\mathrm{lat.}\rangle
		=e^{-i(m_s+m_l+m_\mathrm{lat.})2\pi/5}
		|m_s,m_l,m_\mathrm{lat.}\rangle.
	\end{equation}
	The condition for this state to belong to the $\nu=0$ sector is therefore
	\begin{equation}
		m_s+m_l+m_\mathrm{lat.}\equiv\frac{5}{2}\pmod{5},
	\end{equation}
	or, for $m_s=\pm1/2$,
	\begin{equation}
		m_l+m_\mathrm{lat.}\equiv \pm 2
		\pmod{5}.
		\label{eq:c5-off-axis-condition}
	\end{equation}
	Within the $s$- and $p$-orbital basis, representative combinations satisfying this condition are
	\begin{equation}
		\begin{split}
			&\left|\frac{1}{2},0,2\right\rangle,
			\quad
			\left|\frac{1}{2},1,1\right\rangle,
			\quad
			\left|\frac{1}{2},-1,-2\right\rangle,\\
			&\left|-\frac{1}{2},1,2\right\rangle,
			\quad
			\left|-\frac{1}{2},0,-2\right\rangle,
			\quad
			\left|-\frac{1}{2},-1,-1\right\rangle.
		\end{split}
	\end{equation}
	An additional orbital-species label is suppressed in this notation: the states with $m_l=0$ may be constructed from either an $s$ orbital or the $p_0$ orbital, whereas $m_l=\pm1$ corresponds to the $p_{\pm}$ orbitals. Thus, although axial $s$ and $p$ orbitals cannot enter the $C_5=-1$ sector, off-axis orbitals of both types can do so through the rotational phase associated with their five-site orbits. The protected core mode can therefore be supported by the rings of atoms surrounding the central column. The same five-site orbit structure underlies the block decomposition of the full pentagonal Hamiltonian developed below.
	
	\subsection{Block diagonalization of the pentagonal NW Hamiltonian} \label{sec:block-diag}
	\begin{figure}
		\centering
		\includegraphics[scale=1]{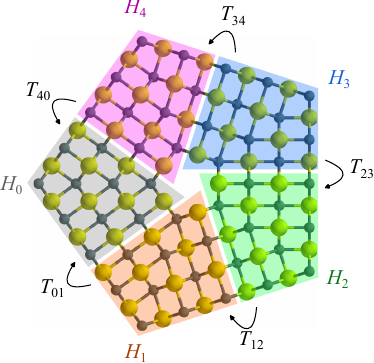}
		\caption{Decomposition of the pentagonal NW, shown in cross section in the $(011)$ plane, into five symmetry-related wedges ordered counterclockwise.}
		\label{fig:pentagon_wedges}
	\end{figure}
	
	We divide the pentagonal NW into five symmetry-related wedges, labeled by the index $n=0,\ldots,4$ and ordered counterclockwise as shown in Fig.~\ref{fig:pentagon_wedges}. For clarity, we first present the block diagonalization for the case in which each wedge couples only to its two nearest neighboring wedges. The inclusion of further-range hopping is straightforward. The Hamiltonian of the NW can be written symbolically as
	\begin{equation}
		\label{eq:formal_hamiltonian}
		H_\mathrm{NW}=
		\begin{pmatrix}
			H_0 & T_{01} & 0 & 0 & T_{40}^{\dagger} \\
			T_{01}^{\dagger} & H_1 & T_{12} & 0 & 0 \\
			0 & T_{12}^{\dagger} & H_2 & T_{23} & 0 \\
			0 & 0 & T_{23}^{\dagger} & H_3 & T_{34} \\
			T_{40} & 0 & 0 & T_{34}^{\dagger} & H_4
		\end{pmatrix}.
	\end{equation}
	Here, $H_n$ contains all onsite and intrawedge terms in wedge $n$, while $T_{n,n+1}$ contains the hopping terms across the boundary between two neighboring wedges. The spin--orbital basis is defined with respect to a common global Cartesian frame. If $U_5$ denotes the unitary representation of an active counterclockwise rotation by $2\pi/5$ acting on the internal spin--orbital degrees of freedom, the five diagonal and interwedge blocks satisfy
	\begin{align}
		H_n&= U_5^n H_0 \left(U_5^\dagger\right)^n,
		\label{eq:Hn_rotation}\\
		T_{n,n+1}&= U_5^n T_{01} \left(U_5^\dagger\right)^n .
		\label{eq:Tn_rotation}
	\end{align}
	For spinful states,
	\begin{equation}
		U_5^5=-\mathbb I.
		\label{eq:U5_fifth}
	\end{equation}
	
	In the wedge basis, the full fivefold rotation is represented by
	\begin{equation}
		\hat C_5=P_5\otimes U_5,
		\label{eq:C5_wedge}
	\end{equation}
	where $P_5$ is the cyclic permutation matrix defined in Eq.~\eqref{eq:c5_offaxis_shift}. Eqs.~\eqref{eq:Hn_rotation} and~\eqref{eq:Tn_rotation} imply $[\hat C_5,H_\mathrm{NW}]=0$. For each rotational eigenvalue $\lambda_\nu$ defined in Eq.~\eqref{eq:c5_eigenvalues}, we introduce the basis matrix
	\begin{equation}
		Q_\nu=
		\frac{1}{\sqrt{5}}
		\begin{pmatrix}
			\mathbb I \\
			\lambda_\nu^*U_5 \\
			\left(\lambda_\nu^*U_5\right)^2 \\
			\left(\lambda_\nu^*U_5\right)^3 \\
			\left(\lambda_\nu^*U_5\right)^4
		\end{pmatrix},
		\label{eq:Qm}
	\end{equation}
	whose columns form an orthonormal basis of the corresponding eigenspace of $\hat C_5$. These matrices satisfy
	\begin{equation}
		\hat C_5 Q_\nu=\lambda_\nu Q_\nu,
		\quad
		Q_\nu^\dagger Q_{\nu'}=\delta_{\nu\nu'}\mathbb I.
	\end{equation}
	
	The unitary matrix
	\begin{equation}
		Q=\begin{pmatrix}Q_{-2}&Q_{-1}&Q_0&Q_1&Q_2\end{pmatrix}
	\end{equation}
	diagonalizes both the rotation operator and the Hamiltonian,
	\begin{align}
		Q^\dagger\hat C_5Q
		&=\bigoplus_{\nu=-2}^{2}\lambda_\nu\mathbb I,\\
		Q^\dagger H_\mathrm{NW}Q
		&=\bigoplus_{\nu=-2}^{2}H_\mathrm{NW}^{\{\nu\}},
	\end{align}
	where
	\begin{equation}
		\begin{split}
			H_\mathrm{NW}^{\{\nu\}}
			&=Q_\nu^\dagger H_\mathrm{NW} Q_\nu\\
			&=H_0+\lambda_\nu^*T_{01}U_5
			+\lambda_\nu U_5^\dagger T_{01}^\dagger.
		\end{split}
		\label{eq:Hm}
	\end{equation}
	Equation~\eqref{eq:Hm} shows that each rotational sector is represented by  an effective single-wedge Hamiltonian whose two lateral boundaries are connected by a sector-dependent twisted hopping.

	\subsection{Relation of the $C_5=-1$ sector to a single-TP shell}
	
\begin{figure}
		\centering
		\includegraphics[scale=1]{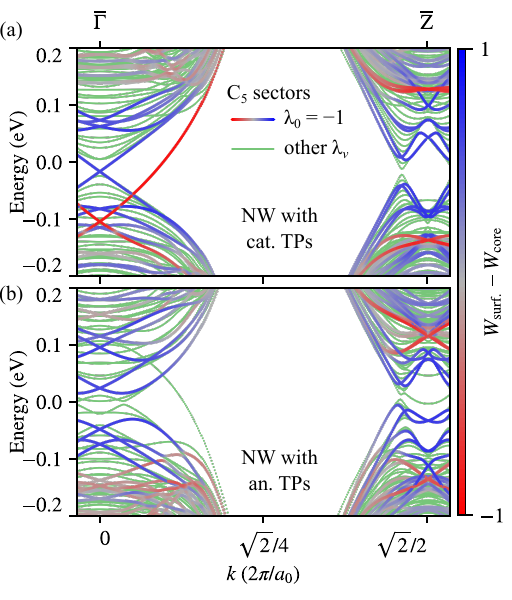}
		\caption{Rotationally resolved band structures of pentagonal SnTe NWs with (a) cationic and (b) anionic TPs. The spectra correspond to those shown in Figs.~\ref{fig:nw_slab_spectra}(a,b). States belonging to the $C_5=-1$ sector are shown with thicker lines and are colored according to their localization near the core or the outer surface, following the convention of Figs.~\ref{fig:nw_slab_spectra}(a,b), while all other rotational sectors are shown in light green. In the cationic wire, the core and outer-surface Dirac crossings near $\Gammabar$ both belong to the $C_5=-1$ sector, whereas no corresponding crossing occurs in this sector for the anionic wire.}
		\label{fig:c5_resolved_nw}
	\end{figure}

	For $\nu=0$, one has $\lambda_0=-1$, and Eq.~\eqref{eq:Hm} becomes
	\begin{equation}
		H_\mathrm{NW}^{\{0\}}=H_0-T_{01}U_5-U_5^\dagger T_{01}^\dagger.
		\label{eq:Hm_zero}
	\end{equation}
	A direct numerical verification of the rotational decomposition is shown in Fig.~\ref{fig:c5_resolved_nw}. The two panels reproduce the cationic- and anionic-TP spectra of Figs.~\ref{fig:nw_slab_spectra}(a,b), but now distinguish the $C_5=-1$ sector from the remaining rotational sectors. In the cationic wire, both the core and outer-surface Dirac crossings near $\Gammabar$ belong to the $\nu=0$ sector. The corresponding sector of the anionic wire contains no such crossing. The distinction between the two TP sublattice types is therefore present within the self-conjugate rotational block.
	
	The correspondence between the $C_5=-1$ block and the single-TP cylindrical shell introduced in Sec.~\ref{sec:cyl_slab_TP} follows from the wedge decomposition shown in Fig.~\ref{fig:pentagon_wedges}. Each wedge is centered on one TP and bounded laterally by regular $(01\bar{1})$ atomic planes. We label the wedges so that wedge $0$ has the same lattice orientation as the single-TP shell considered in the main text. The relation between the two constructions is then determined by the hopping that reconnects the two lateral boundaries of this wedge. In the single-TP shell, this role is played by the seam hopping shown in Fig.~\ref{fig:TP_slab_spectra}(c).
	
	Following the prescription of Ref.~\cite{Geier2021} for hopping across a Volterra branch cut, we write the shell Hamiltonian as
	\begin{equation}
		H_{\mathrm{shell}}
		=
		H_{0,\mathrm{shell}}+H_{\mathrm{seam}},
		\label{eq:shell_hamiltonian_decomposition}
	\end{equation}
	where $H_{0,\mathrm{shell}}$ describes the shell with the two sides of the cut disconnected. The seam contribution takes the form
	\begin{equation}
		H_{\mathrm{seam}}
		=
		t_{\mathrm{seam}}
		\otimes U(\alpha)H_{\mathrm{hop}}
		+\mathrm{h.c.},
		\label{eq:shell_seam_hamiltonian}
	\end{equation}
	where $t_{\mathrm{seam}}$ connects the two sets of lattice sites across the seam, $U(\alpha)$ is the active spin--orbital rotation relating the local coordinate frames on the two sides of the seam, and $H_{\mathrm{hop}}$ denotes the corresponding regular hopping matrix in spin--orbital space. Thus, $t_{\mathrm{seam}}$ acts only in the lattice-site space, whereas $U(\alpha)H_{\mathrm{hop}}$ acts in the spin--orbital space.
	
	The relation to the reduced pentagonal Hamiltonian follows by comparing the wedge geometries in Figs.~\ref{fig:TP_slab_spectra}(c) and \ref{fig:pentagon_wedges}. With the boundary orbitals ordered consistently in the two constructions,
	\begin{equation}
		H_{0,\mathrm{shell}}\simeq H_0,
		\qquad
		T_{01}^\dagger
		\simeq
		t_{\mathrm{seam}}\otimes H_{\mathrm{hop}}.
		\label{eq:shell_wedge_identification}
	\end{equation}
	The approximations account for the weak deformation required to convert the natural crystallographic wedge angle into the exact fivefold angle of the pentagonal NW.
	
	Indeed, Eq.~\eqref{eq:single_tp_alpha} gives
	\begin{equation}
		\alpha
		=
		2\pi-\arccos(1/3)
		\simeq
		2\pi-\frac{2\pi}{5}.
	\end{equation}
	Using $U_5=U(2\pi/5)$ and the spinful relation $U(2\pi)=-\mathbb I$, one therefore obtains
	\begin{equation}
		\begin{split}
			U(\alpha)
			&\simeq
			U\left(2\pi-\frac{2\pi}{5}\right)\\
			&=
			U(2\pi)U^\dagger\left(\frac{2\pi}{5}\right)
			=
			-U_5^\dagger.
		\end{split}
		\label{eq:shell_rotation_c5_relation}
	\end{equation}
	Equations~\eqref{eq:shell_seam_hamiltonian}--\eqref{eq:shell_rotation_c5_relation} then give
	\begin{equation}
		H_{\mathrm{seam}}
		\simeq
		-U_5^\dagger T_{01}^\dagger
		-T_{01}U_5,
		\label{eq:shell_seam_c5}
	\end{equation}
	where the rotation operators are implicitly extended by the identity in the lattice-site space. Consequently,
	\begin{equation}
		\begin{split}
			H_{\mathrm{shell}}
			&\simeq
			H_0-T_{01}U_5-U_5^\dagger T_{01}^\dagger\\
			&=
			H_\mathrm{NW}^{\{0\}}.
		\end{split}
		\label{eq:shell_c5_block_equivalence}
	\end{equation}
	
	Equation~\eqref{eq:shell_c5_block_equivalence} establishes that, up to the weak strain deformation, the single-TP shell realizes the complete $C_5=-1$ block of the pentagonal NW Hamiltonian. The inner- and outer-boundary Dirac crossings of the cationic shell in Fig.~\ref{fig:TP_slab_spectra}(a) therefore correspond to the core and outer-surface crossings highlighted in Fig.~\ref{fig:c5_resolved_nw}(a). Likewise, the absence of analogous crossings in the anionic shell, shown in Fig.~\ref{fig:TP_slab_spectra}(b), agrees with the gapped $C_5=-1$ sector of the anionic NW in Fig.~\ref{fig:c5_resolved_nw}(b). The remaining quantitative differences between the corresponding spectra arise predominantly from the smaller thickness of the single-TP shell.
	
	The seam-stitching construction used above for the $C_5=-1$ block also generates the remaining $C_5$ sectors. Multiplying the seam rotation by a sector-dependent phase,
	\begin{equation}
		U_\nu(\alpha)=-\lambda_\nu U(\alpha),
	\end{equation}
	gives
	\begin{equation}
		H_{\mathrm{shell}}^{\{\nu\}}
		\simeq
		H_0+\lambda_\nu^*T_{01}U_5
		+\lambda_\nu U_5^\dagger T_{01}^\dagger
		=
		H^{\{\nu\}}_\mathrm{NW}.
	\end{equation}
	Thus, every rotational block of the pentagonal NW can be represented by a single-TP shell with a suitably modified seam hopping.
	
	More generally, the construction can be applied to idealized NWs obtained by changing the number of symmetry-related wedges. For example, removing one wedge from the pentagonal NW and stitching the two exposed boundaries produces a four-TP geometry with $C_4$ symmetry, whose rotational sectors can be analyzed by the same single-wedge construction. This procedure specifies the corresponding boundary hopping but does not account for the strain generated when the remaining crystal is deformed to close the NW. For spinful rotations, a sector with $C_n$ eigenvalue $-1$ exists only when $n$ is odd. In that case, the corresponding self-conjugate rotational block has the same form as the $C_5=-1$ Hamiltonian derived above. For even $n$, no $C_n=-1$ sector exists, all rotational sectors occur in complex-conjugate pairs, and the symmetry-protected crossing associated with the self-conjugate block is therefore absent.
	
	\subsection{Projected symmetries of the five-channel model}
	\label{app:kp_symmetries}
	
	We finally derive the symmetry matrices used in the $k\cdot p$ theory near $\Gammabar$. The basis consists of five helical Kramers pairs ordered counterclockwise around the NW, with a two-component spinor attached to each channel. The labeling is chosen so that $C_5$ maps channel $n$ onto $n+1$ and channel $n=0$ lies in the mirror plane $M_y$. In the real-space channel basis, a convenient representation is
	\begin{equation}
		\begin{split}
			\hat C_5 &=P_5\otimes u_5,
			\qquad u_5=\exp\left(-\frac{i\pi}{5}s_z\right),\\
			\hat M_y &=R_y\otimes(-is_y),\\
			\hat M_z &=\mathbb I_{5\times5}\otimes(-is_z),\\
			\hat\Theta &=\mathbb I_{5\times5}\otimes is_yK,
		\end{split}
		\label{eq:kp_symmetry_real_space}
	\end{equation}
	where $P_5$ is given in Eq.~\eqref{eq:c5_offaxis_shift}, $u_5$ is the low-energy representation of the microscopic spin--orbital rotation $U_5$ introduced in Section~\ref{sec:block-diag}, and
	\begin{equation}
		R_y=
		\begin{pmatrix}
			1&0&0&0&0\\
			0&0&0&0&1\\
			0&0&0&1&0\\
			0&0&1&0&0\\
			0&1&0&0&0
		\end{pmatrix}
	\end{equation}
	implements the reflection with respect to the $M_y$ plane. The rotational decomposition derived in Section~\ref{sec:block-diag} applies directly to the five-channel model. In Eq.~\eqref{eq:Qm}, the microscopic spin--orbital rotation $U_5$ is replaced by $u_5$, and the resulting matrices $Q_\nu$ project the channel basis onto the rotational sectors:
	\begin{align}
		Q_\nu^\dagger\hat C_5Q_{\nu'}
		&=\lambda_\nu  \mathbb{I}_{2\times2}  \,\delta_{\nu,\nu'},\\
		Q_\nu^\dagger\hat M_zQ_{\nu'}
		&=-is_z\,\delta_{\nu,\nu'},\\
		Q_\nu^\dagger\hat M_yQ_{\nu'}
		&=-is_y\,\delta_{\nu',-\nu},\\
		Q_\nu^\dagger\hat\Theta Q_{\nu'}
		&=is_y\,\delta_{\nu',-\nu}K.
	\end{align}
	Thus, $C_5$ and $M_z$ preserve each rotational sector, whereas $M_y$ and time reversal exchange the sectors $\nu$ and $-\nu$.
	
	In the self-conjugate $\nu=0$ sector, all symmetries act within a single two-dimensional space and take the form
	\begin{equation}
		\begin{split}
			C_5^{\{0\}}&=-\mathbb{I}_{2\times2},
			\qquad M_y^{\{0\}}=-is_y,\\
			M_z^{\{0\}}&=-is_z,
			\qquad \Theta^{\{0\}}=is_yK.
		\end{split}
		\label{eq:app_gamma_symmetry_minus_one}
	\end{equation}
	For $\nu=1,2$, the ordered basis $(Q_\nu,Q_{-\nu})$ gives
	\begin{equation}
		\begin{split}
			C_5^{\{\nu,-\nu\}}&=
			\begin{pmatrix}
				\lambda_\nu \mathbb{I}_{2\times2} &0\\
				0&\lambda_{-\nu} \mathbb{I}_{2\times2}
			\end{pmatrix},\\
			M_y^{\{\nu,-\nu\}}&=
			\begin{pmatrix}
				0&-is_y\\
				-is_y&0
			\end{pmatrix},\\
			M_z^{\{\nu,-\nu\}}&=
			\begin{pmatrix}
				-is_z&0\\
				0&-is_z
			\end{pmatrix},\\
			\Theta^{\{\nu,-\nu\}}&=
			\begin{pmatrix}
				0&is_y\\
				is_y&0
			\end{pmatrix}K.
		\end{split}
		\label{eq:app_gamma_symmetry_paired}
	\end{equation}
	Equations~\eqref{eq:app_gamma_symmetry_minus_one} and~\eqref{eq:app_gamma_symmetry_paired} thus provide the derivation of the symmetry representations used in Eqs.~\eqref{eq:gamma_symmetry_minus_one} and~\eqref{eq:gamma_symmetry_paired}.
	
	\section{Closed-surface quantization and hybridized mirror modes near $\Zbar$}
	\label{app:zbar}
	This appendix explains why the spectra near $\Zbar$ remain gapped for both TP sublattice types using complementary momentum- and real-space descriptions. We first relate the NW spectrum near $\Zbar$ to the quantized spectrum of a wrapped SnTe $(100)$ surface. We then resolve the same spectrum into ten mirror-related helical channels on each boundary, with five localized at the TP edges and five at the face midlines. This channel decomposition establishes the connection to the mirror-protected side-surface states of the twinning superlattices studied in Ref.~\cite{Samadi2022_TP} and clarifies why the corresponding modes can instead hybridize into a gapped spectrum on the closed NW surface. We show how the individual channels can be exposed by mirror-preserving surface perturbations and describe their hybridization using a low-energy $k\cdot p$ theory. A geometric surface-domain-wall analysis then shows how the wrapped boundary condition permits the TP-edge and face-midline modes to hybridize and open a gap. Both the momentum-space quantization and real-space channel constructions apply to the outer and core boundaries, but the core perimeter produces a much larger level spacing and gap. Finally, we confirm the persistence of the $\Zbar$ gaps in the material-specific multiorbital calculations.
	
	\subsection{Closed-surface quantization}
	\label{sec:slab-approx}
	
\begin{figure*}
		\centering
		\includegraphics[scale=1]{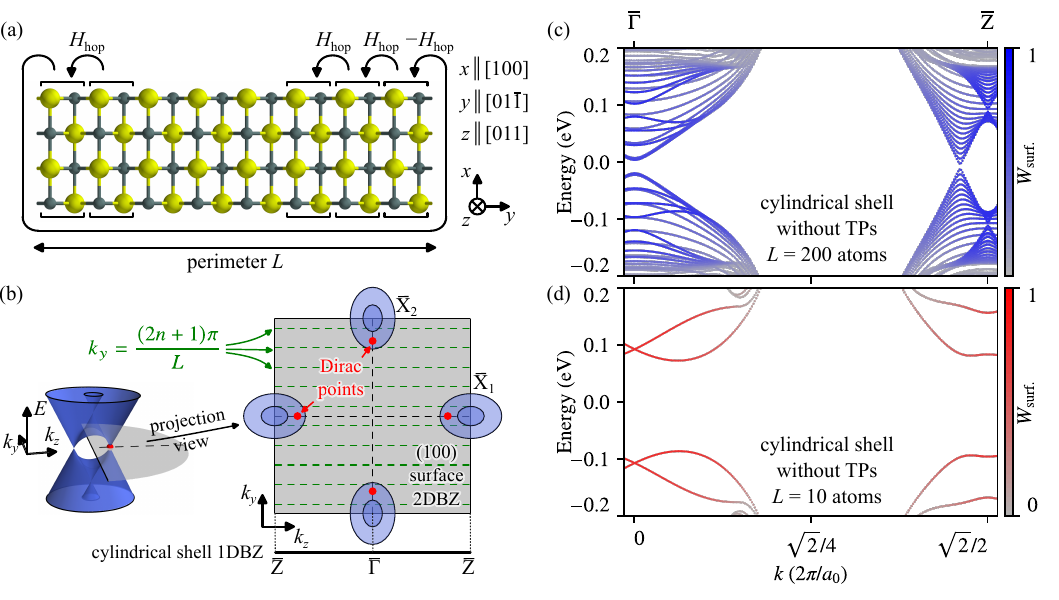}
		\caption{Closed-surface origin of the gap near $\Zbar$. (a) Construction of the cylindrical shell from a locally flat $(100)$-oriented slab, illustrated for a shell four atomic layers thick with $L=16$ atoms. (b) Projection of the extended $(100)$ surface spectrum onto the one-dimensional shell Brillouin zone. The surface band structure is shown schematically in blue, and the green dashed lines denote the transverse momenta $k_y=(2n+1)\pi/L$. (c,d) Spectra of defect-free cylindrical shells. Panel (c) shows a shell with transverse perimeter $L=200$ atoms, matching the NW perimeter in Fig.~\ref{fig:nw_slab_spectra}, while panel (d) shows a shell with $L=10$ atoms, chosen to match the perimeter of the innermost atomic ring in the NW. Both shells are 20 atomic layers thick. The shell-localized states are shown in blue in panel (c) and red in panel (d), following the color convention used for the outer-surface and core states, respectively, in Fig.~\ref{fig:nw_slab_spectra}.}
		\label{fig:zbar_quantization}
	\end{figure*}

	We begin with a defect-free analogue of the single-TP shell, obtained by wrapping a SnTe $(100)$ slab around the $[011]$ axis to form a cylindrical shell. As shown schematically in Fig.~\ref{fig:zbar_quantization}(a), the local hopping amplitudes and onsite terms are kept identical to those of the corresponding unwrapped geometry, while the two ends of the transverse direction are connected by a rotated boundary hopping. In the absence of a TP, the accumulated rotation around the shell perimeter is $2\pi$, so that $U(2\pi)=-\mathbb I$ in the spinful basis. The seam hopping therefore differs from the regular hopping by a minus sign, and the wave function obeys antiperiodic boundary conditions around the perimeter.
	
	If $L$ denotes the perimeter and $y$ is the wrapped direction, the allowed transverse momenta are
	\begin{equation}
		k_y=\frac{(2n+1)\pi}{L},
		\qquad n\in\mathbb Z.
		\label{eq:zbar_antiperiodic_quantization}
	\end{equation}
	The one-dimensional shell spectrum is consequently obtained from discrete momentum cuts through the band structure of an extended $(100)$ slab, as illustrated in Fig.~\ref{fig:zbar_quantization}(b). For a shell sufficiently thick in the $x$ direction indicated in Fig.~\ref{fig:zbar_quantization}(a), each boundary approaches the known SnTe $(100)$ surface spectrum, whose four Dirac points occur at $(k_y,k_z)=(0,\pm k_\Lambda)$ and $(\pm k_\Lambda,0)$~\cite{hsieh2012topological,Liu2013}. The Dirac points projected near $\Zbar$ lie on the $k_y=0$ cut, which is excluded by Eq.~\eqref{eq:zbar_antiperiodic_quantization}. The closed surface is therefore gapped, with a confinement scale expected to decrease as $1/L$.
	
	Within a given shell calculation, the inner and outer boundaries have the same transverse period $L$ and therefore the same confinement scale. The outer-surface and core spectra of the pentagonal NW are instead represented by the separate shell calculations shown in Figs.~\ref{fig:zbar_quantization}(c) and \ref{fig:zbar_quantization}(d), respectively. Both shells have a radial wall thickness of 20 atomic layers, matching that used in the NW calculations of Fig.~\ref{fig:nw_slab_spectra}. The large-perimeter shell in panel (c), with $L=200$ atoms, matches the NW perimeter and corresponds to the extended outer NW surface. Its surface-localized states are shown in blue, following the outer-surface color convention of Fig.~\ref{fig:nw_slab_spectra}. The small-perimeter shell in panel (d), with $L=10$ atoms, matches the perimeter of the innermost atomic ring in the NW and corresponds to the NW core region. Its surface-localized states are shown in red, following the corresponding core-state convention. The smaller perimeter produces a larger confinement gap and a wider spacing between the one-dimensional subbands. Finite radial thickness produces quantitative deviations from the ideal surface-state spectrum but does not alter the confinement mechanism associated with quantization around the shell perimeter. 
	
	Furthermore, we note that the antiperiodic boundary condition in the cylindrical shell explains the difference from the extended side surfaces studied in Ref.~\cite{Samadi2022_TP}. A periodically repeated flat surface admits transverse momenta $k_y=2n\pi/L$ and therefore contains the $k_y=0$ sector. The mirror-protected $\Mbar$ crossings found for the side surface of a twinning superlattice and the gapped $\Zbar$ spectrum of the closed NW thus may be understood as arising from different transverse boundary conditions. 
	
	Introducing a TP, as in the single-TP shell of Sec.~\ref{sec:cyl_slab_TP}, locally perturbs the wrapped surface spectrum. Near $\Zbar$, the TP acts as a local perturbation, modifies the extended surface states and generates trivial shallow Rashba-split subbands near the TP--surface junction. The five TPs of the complete pentagonal NW produce analogous local modifications at their surface terminations while preserving the same closed-surface quantization mechanism.
	
	\subsection{Mirror-related TP-edge and face-midline channels}
\begin{figure}
		\centering
		\includegraphics[scale=1]{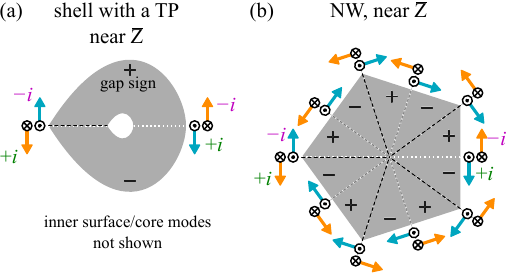}
		\caption{Mirror-related helical channels near $\Zbar$. (a) Schematic surface-gap pattern for a cylindrical shell containing a single TP. The mirror-breaking surface mass changes sign across the TP mirror plane and the opposite regular $(01\bar{1})$ mirror plane, producing a helical Kramers pair at each domain wall. (b) Corresponding pattern for the pentagonal NW, where the mass changes sign across the $M_y$ plane and its $C_5$ related counterparts, and produces channels at the TP edges and at the midlines of the opposite faces. The colored arrows indicate the local spin polarizations perpendicular to the corresponding mirror planes, while the circle-dot and circle-cross symbols denote opposite propagation directions along the NW axis. Green and magenta label the $M_y=+i$ and $M_y=-i$ sectors, respectively. Only the outer-boundary channels are shown; the corresponding inner-boundary or core channels are omitted for clarity.}
		\label{fig:zbar_domain_walls}
	\end{figure}

	The same gapped spectrum admits a complementary real-space interpretation. The cylindrical shell containing one TP has two mirror planes relevant near $\Zbar$: the mirror plane coinciding with the TP and a regular $(01\bar{1})$ mirror plane on the opposite side of the shell. Consider a time-reversal-preserving surface perturbation that locally breaks the $(01\bar{1})$ mirror symmetry of the $(100)$ surface away from these two planes, while preserving the TP mirror plane and the opposite regular mirror plane. The associated surface mass is odd under reflection across either preserved plane and must therefore change sign across it. Their intersections with a given surface consequently form mirror-symmetric mass domain walls, as shown schematically in Fig.~\ref{fig:zbar_domain_walls}(a). According to the hinge mechanism of Ref.~\cite{schindler2018higher}, each domain wall supports a helical Kramers pair near $\Zbar$. The outer surface of the shell therefore carries one pair at the TP termination and another at the termination of the regular mirror plane, with the same two-pair structure occurring on the inner surface. Once the two pairs on a given boundary are spatially separated, each isolated Kramers pair remains gapless under time-reversal symmetry even if mirror symmetry is weakly broken locally. When the surface-gapping perturbation is removed, however, the intervening surface becomes conducting and the two pairs can hybridize through the extended surface states.
	
	A weak, spatially varying realization of this surface mass is already implicit in the single-TP shell and in the pentagonal geometry. In the pentagonal NW, each face away from a TP edge locally resembles a flat $(100)$ surface and retains an approximate local $(01\bar{1})$ mirror symmetry which is exact on the face midline, while the atomic environment becomes increasingly asymmetric on approaching the TP edge. However, the TP itself introduces a distinct local mirror symmetry. The geometry can therefore be viewed as generating a nonuniform surface mass that, by symmetry, changes sign across the TP edges and the midlines of the faces. The corresponding mass pattern on the full NW is shown in Fig.~\ref{fig:zbar_domain_walls}(b), together with the global $M_y$ eigenvalues and local spin polarizations of the associated domain-wall modes. Because the intrinsic mass remains weak over much of each face, these modes are broadened into extended surface states rather than appearing as spectrally isolated channels. The momentum-space closed-surface quantization and the real-space domain-wall construction thus provide complementary descriptions of the same gapped spectrum.
	
	In the NW, each of the five $C_5$-related mirror planes carries one TP-edge pair and one face-midline pair on the outer NW surface. When spectrally isolated, the outer surface therefore supports five TP-edge helical pairs and five face-midline pairs near $\Zbar$. The same channel decomposition applies to an inner boundary and, by continuity, to the microscopic core region obtained by shrinking that boundary. As the boundary is contracted, the ten helical pairs are brought into close proximity and hybridize more strongly, producing a much larger gap than on the extended outer surface. 
	
	This channel counting also clarifies the relation to the mirror Chern numbers obtained for the twinning superlattices in Ref.~\cite{Samadi2022_TP}. That work assigned mirror Chern numbers of magnitude $2$ and $1$ to cationic and anionic TPs, respectively. Consistently, each cationic TP edge in the NW supports two helical pairs distributed between the two momentum regions: one near $\Gammabar$ and one near $\Zbar$. An anionic TP edge supports only the pair near $\Zbar$, which is the NW counterpart of the mirror-protected crossing near $\Mbar$ in the extended superlattice side-surface geometry.
	
	The closed NW and the single-TP shell contain, in addition, regular mirror planes that are absent in the twinning superlattices. Their intersections with the boundary generate an additional helical pair at the face midline opposite each TP edge. The complete NW with cationic TPs therefore contains five TP-edge pairs near $\Gammabar$, five TP-edge pairs near $\Zbar$, and five face-midline pairs near $\Zbar$, giving fifteen pairs in total. For anionic TPs, only the five TP-edge and five face-midline pairs near $\Zbar$ remain, giving ten pairs. In particular, the $\Zbar$ sector contains an even number of pairs for either TP type, so channel parity alone does not require a Dirac crossing there. Although crystalline symmetry could in principle protect such a crossing, the calculated NW spectra remain gapped near $\Zbar$. The following subsections show that, in the wrapped geometry, the TP-edge and face-midline modes hybridize within the symmetry sectors required to open this gap.
	
	The modes underlying this counting can be separated from the extended surface states by applying stronger perturbations to the outermost atomic layer. The perturbation used in Fig.~\ref{fig:pert_NW_spectra}(a,b) breaks the five mirror symmetries represented by $M_y$ and its $C_5$-related counterparts. It gaps the extended TCI surface states without producing isolated channels near $\Zbar$ for either TP sublattice type.
	
\begin{figure}
		\centering
		\includegraphics[scale=1]{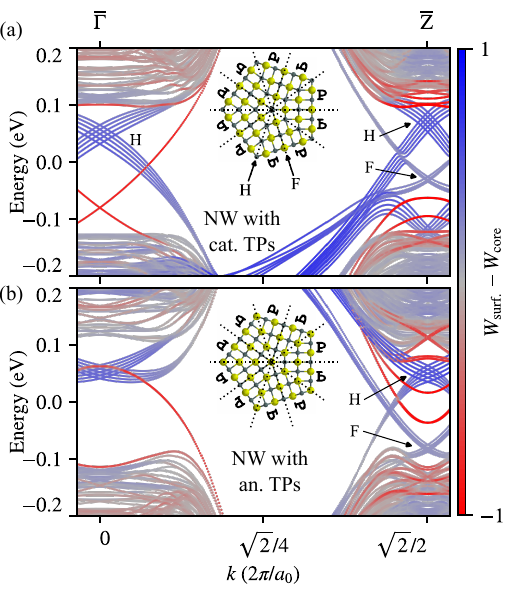}
		\caption{Band structures of pentagonal SnTe NWs after applying mirror- and time-reversal-preserving perturbations to the outermost atomic layers. Panels (a) and (b) correspond to NWs with cationic and anionic TPs, respectively, with the same geometries as in Fig.~\ref{fig:nw_slab_spectra}. The inset schematics indicate the perturbed surface regions, with $P$ denoting mirror-breaking perturbations arranged so that each $M_y$ symmetry is preserved at a TP edge and at the midline of the opposite face. Near $\Zbar$, both TP types exhibit five TP-edge pairs, labeled H, and five face-midline pairs, labeled F.}
		\label{fig:zbar_channels}
	\end{figure}

	To expose the domain-wall channels, we instead choose a perturbation that gaps the surface away from the mirror lines while preserving each $M_y$ symmetry at the corresponding TP edge and opposite face midline. The perturbation varies across every face such that the local perturbation Hamiltonians on opposite sides of the face midline are mapped onto one another by $M_y$. The same pattern is repeated on all five faces, as indicated by the inset schematics in Fig.~\ref{fig:zbar_channels}(a,b). Weak TP-edge- and midline-dependent electrostatic potentials are then added to separate the otherwise degenerate channels in energy. Although these potentials formally break global $C_5$ symmetry and may weakly break $M_y$ away from a given channel, they remain locally mirror symmetric where the corresponding wave functions are localized and therefore do not appreciably mix the two mirror eigenspaces.
	
	As shown in Fig.~\ref{fig:zbar_channels}, the resulting spectra contain ten helical Kramers pairs near $\Zbar$ for both TP sublattice types: five localized at the TP edges and five at the face midlines. Near $\Gammabar$, the same perturbation exposes five additional TP-edge pairs only in the cationic-TP NW, in agreement with the channel counting above. Removing the artificial perturbation restores the conducting faces and allows the $\Zbar$ channels to hybridize around the closed perimeter.
	
	\subsection{Doubled low-energy theory}
	\label{sec:doubled-kp}
	
	The hybridization of the ten channels near $\Zbar$ is described by a doubled version of the five-pair model introduced for the $\Gammabar$ sector in Eqs.~\eqref{eq:gamma_symmetry_minus_one}--\eqref{eq:gamma_parameter_constraints}. The two copies represent the TP-edge and face-midline modes, respectively. Since both sets of channels transform identically under $C_5$, $M_y$, $M_z$, and $\Theta$, each symmetry $S\in\{C_5,M_y,M_z,\Theta\}$ has the doubled representation
	\begin{equation}
		\widetilde S=
		\begin{pmatrix}
			S&0\\
			0&S
		\end{pmatrix}.
		\label{eq:zbar_doubled_symmetry}
	\end{equation}
	The rotational sectors are labeled by the same eigenvalues $\lambda_\nu$ as in Eq.~\eqref{eq:c5_eigenvalues}. In a given sector, the most general Hamiltonian through first order in the momentum $k_z$ measured from $\Zbar$ has the block form
	\begin{equation}
		\widetilde H^{\{\nu\}}=
		\begin{pmatrix}
			H_h^{\{\nu\}}&C^{\{\nu\}}\\
			{C^{\{\nu\}}}^{\dagger}&H_f^{\{\nu\}}
		\end{pmatrix},
		\label{eq:hinge_face_effective}
	\end{equation}
	where
	\begin{equation}
		H_a^{\{\nu\}}=
		\epsilon_{a,\nu}s_0+
		\Delta_{a,\nu}s_z+
		v_{a,\nu}k_zs_y,
		\label{eq:zbar_diagonal_blocks}
	\end{equation}
	where $a\in\{h,f\}$ with $h$ denoting the TP-edge modes and $f$ the face-midline modes, and
	\begin{equation}
		C^{\{\nu\}}=
		t_\nu s_0+
		t'_\nu s_z+
		\left(u_\nu s_y+iu'_\nu s_x\right)k_z.
		\label{eq:zbar_coupling_block}
	\end{equation}
	Time reversal and the crystalline symmetries impose
	\begin{equation}
		\begin{alignedat}{2}
			\epsilon_{h,\nu}&=\epsilon_{h,-\nu},\qquad
			&\epsilon_{f,\nu}&=\epsilon_{f,-\nu},\\
			\Delta_{h,\nu}&=-\Delta_{h,-\nu},
			&\Delta_{f,\nu}&=-\Delta_{f,-\nu},\\
			v_{h,\nu}&=v_{h,-\nu},
			&v_{f,\nu}&=v_{f,-\nu},\\
			t_\nu&=t_{-\nu},
			&t'_\nu&=-t'_{-\nu},\\
			u_\nu&=u_{-\nu},
			&u'_\nu&=-u'_{-\nu}.
		\end{alignedat}
		\label{eq:10_constraints}
	\end{equation}
	
\begin{figure}
		\centering
		\includegraphics[scale=1]{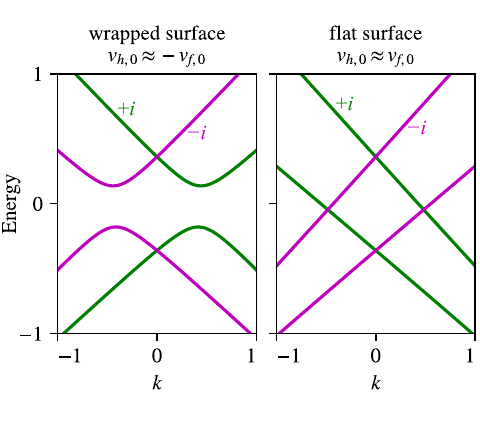}
		\caption{Mirror-resolved spectra of the doubled low-energy model in the self-conjugate sector $\nu=0$ (the remaining rotational sectors are not shown). The spectra describe a single boundary, corresponding either to the outer surface or to the core boundary. (a) Wrapped geometry with $v_{h,0}\simeq-v_{f,0}$. Counterpropagating TP-edge and face-midline branches belong to the same $M_y$ eigenspace and hybridize, producing an anticrossing. (b) Flat geometry with $v_{h,0}\simeq v_{f,0}$. The counterpropagating branches belong to opposite mirror eigenspaces and remain decoupled. Green and magenta denote the $M_y=+i$ and $M_y=-i$ sectors, respectively, following the color convention of Fig.~\ref{fig:zbar_domain_walls}.}
		\label{fig:zbar_kp}
	\end{figure}

	The sectors $\nu=\pm1$ and $\nu=\pm2$ are generally massive, so the possible low-energy crossing is controlled by the self-conjugate sector $\nu=0$. The corresponding mirror-resolved spectra are illustrated in Fig.~\ref{fig:zbar_kp}. When $v_{h,0}\simeq v_{f,0}$, the counterpropagating branches belong to opposite $M_y$ eigenspaces and cannot hybridize, as shown in Fig.~\ref{fig:zbar_kp}(b). When $v_{h,0}\simeq-v_{f,0}$, counterpropagating TP-edge and face-midline branches occur within the same $M_y$ eigenspace, and the symmetry-allowed coupling opens an anticrossing, as shown in Fig.~\ref{fig:zbar_kp}(a). The latter configuration is realized on the closed boundaries of the pentagonal NW. The following subsection provides a geometric interpretation of the opposite velocity signs by relating the cylindrical-shell, single-TP-shell, and pentagonal-NW geometries, and connects this picture to an auxiliary $C_2$ limit of the doubled low-energy model. 
	
	\subsection{Geometric interpretation of the opposite velocities}
	\label{sec:zbar_wrapped_chirality}
	
\begin{figure*}[!htbp]
		\centering
		\includegraphics[scale=1]{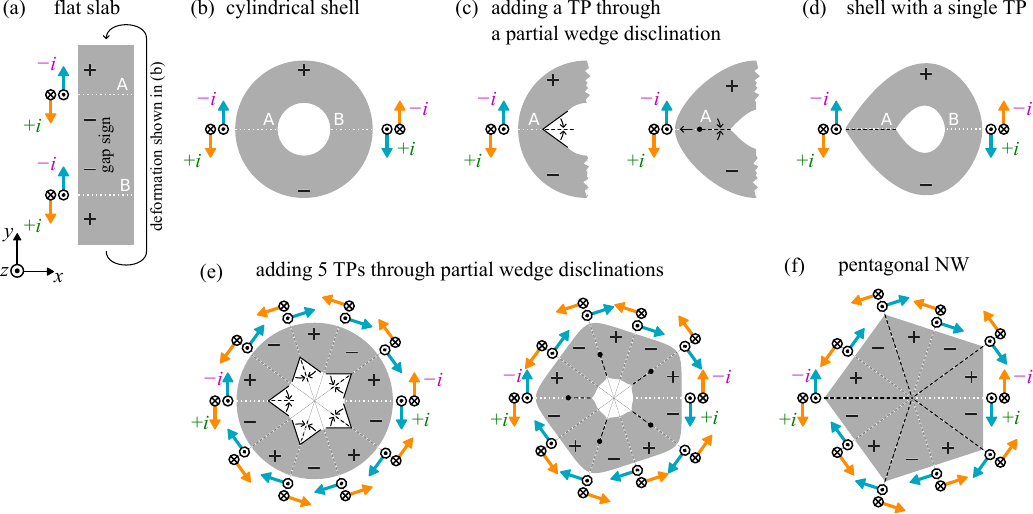}
		\caption{Geometric interpretation of the mirror-resolved chiralities near $\Zbar$. All panels show cross sections in the $(011)$ plane, with the domain-wall modes propagating along the $[011]$ direction perpendicular to the page. Only the outer-boundary channels are shown, except in panel (a), where only the corresponding channels on the left-hand boundary are displayed. (a) Flat $(100)$-oriented slab truncated by two $(01\bar{1})$ side facets. A mirror-breaking surface mass changes sign across the parallel $(01\bar{1})$ mirror planes A and B, producing a helical Kramers pair at each domain wall. Modes with the same mirror eigenvalue have the same chirality at the two interfaces. The curved arrow indicates the deformation that brings the side facets together to form the closed shell shown in panel (b). (b) Cylindrical shell obtained by wrapping the slab while retaining the mass profile. Plane A remains fixed, whereas the surface at B is rotated by $\pi$. The mode velocities are unchanged, but the spin polarizations at B are reversed in the global coordinate system, so modes with the same chirality at A and B acquire opposite global mirror labels. (c,d) Conversion of plane A into a TP through a partial wedge-disclination construction, yielding a shell with a single TP at A and a regular mirror plane at B. (e,f) Corresponding fivefold construction leading to the pentagonal NW. The signs $+$ and $-$ denote the surface mass, the colored arrows indicate the local spin polarizations, and the circle-dot and circle-cross symbols denote opposite propagation directions along the $[011]$ axis. Green and magenta label the $M_y=+i$ and $M_y=-i$ sectors, respectively.}
		\label{fig:zbar_geometric_construction}
	\end{figure*}

	The relative signs of the TP-edge and face-midline velocities can be visualized through the geometric construction shown in Fig.~\ref{fig:zbar_geometric_construction}. All panels show cross sections in the $(011)$ plane, while the domain-wall modes propagate along the $[011]$ direction perpendicular to the page. We begin with a finite $(100)$-oriented slab truncated laterally by two $(01\bar{1})$ side facets. On its $(100)$ surfaces, we introduce a mirror-breaking mass profile that changes sign across two parallel $(01\bar{1})$ mirror planes, labeled A and B in Fig.~\ref{fig:zbar_geometric_construction}(a), where $x$, $y$, and $z$ correspond to [100], $[01\bar{1}]$, and $[011]$, respectively. The two sign changes form mirror-symmetric domain walls and therefore bind helical Kramers pairs near $\Zbar$~\cite{schindler2018higher}.
	
	In the flat slab, the two domain walls have the same mirror-resolved chirality: modes with the same mirror eigenvalue propagate in the same direction at A and B. Here, chirality denotes the sign of the group velocity along $[011]$ within a fixed mirror eigenspace. The corresponding schematic spectrum is shown in Fig.~\ref{fig:zbar_kp}(b).
	
	This pattern of modes follows from the clean extended $(100)$ surface spectrum of SnTe near $\overline{\mathrm{X}}_1$. As shown in Fig.~\ref{fig:zbar_quantization}(b), this point projects onto $\Zbar$ in the one-dimensional NW Brillouin zone. At $k_y=0$, the surface spectrum contains two co-propagating $M_y=+i$ branches and two oppositely propagating $M_y=-i$ branches~\cite{hsieh2012topological,Liu2013}. Upon introducing an alternating surface-mass profile, the bound modes that form at the mirror-symmetric domain walls inherit the propagation directions of the corresponding mirror sectors of the unperturbed surface. These directions are fixed by the mirror-resolved spectral flow imposed by the bulk TCI topology (encoded in the bulk mirror Chern number) rather than by the sign of the local mass gradient. The modes at A and B therefore have the same chirality within each mirror eigenspace, even though the mass changes in opposite directions across the two domain walls. This local chirality persists upon truncating the system to a finite slab, provided that the domain-wall modes remain spatially separated.
	
	We next wrap the slab around the $z$ direction, parallel to $[011]$, by bringing its two $(01\bar{1})$ side facets together to form the cylindrical shell shown in Fig.~\ref{fig:zbar_geometric_construction}(b), transporting the surface-gap profile together with the surface. This is the same wrapped geometry used in Sec.~\ref{sec:slab-approx}, except that the auxiliary mass profile distinguishes the two mirror planes. During the wrapping, plane A may be held fixed, while the surface at plane B is rotated by $\pi$ about the $[011]$ axis. Because this axis is parallel to the domain walls, the rotation does not reverse the propagation direction of the modes at B. It does, however, reverse their spin polarization in the global coordinate system. The two initially parallel mirror planes become two parts of the same global mirror plane, and the reversed spin polarization at B implies that modes at A and B with the same chirality belong to opposite global $M_y=\pm i$ eigenspaces. Thus, modes belonging to the same global mirror eigenspace have opposite chiralities at A and B in the wrapped shell. The opposite-velocity relation therefore results from the opposite mirror-sector assignments at the two domain walls rather than from a reversal of the local mode velocity.
	
	Plane A can subsequently be converted into a TP through the partial wedge-disclination construction illustrated in Fig.~\ref{fig:zbar_geometric_construction}(c). A wedge with opening angle $\arccos(1/3)$ is removed from the inner side of the shell, exposing two $(111)$ facets. Bending these facets toward A and stitching them together produces a partial wedge disclination of the type discussed in Ref.~\cite{deWit1972}. Increasing the wedge size and moving its apex toward the outer surface continuously transforms the cylindrical shell into the single-TP shell shown in Fig.~\ref{fig:zbar_geometric_construction}(d). Within the $\Zbar$ sector, this construction preserves the chirality of the outer-boundary modes at A. The final shell therefore inherits the wrapped-cylinder pattern: modes in the same global mirror eigenspace have opposite velocities at the TP termination A and at the regular-plane termination B. The corresponding schematic energy spectrum, in which the counterpropagating branches within each mirror eigenspace hybridize and open an anticrossing, is shown in Fig.~\ref{fig:zbar_kp}(a).
	
	At this stage, the imposed surface-gap profile need no longer be regarded as an external perturbation. The single-TP geometry itself breaks the local $(01\bar{1})$ mirror symmetry away from A and B and thus realizes the weak, spatially varying surface mass discussed in the preceding subsection. In the unperturbed shell this mass is too weak to isolate the domain-wall modes from the extended surface spectrum, but the mirror-resolved channel assignment remains the same.
	
	The same construction can be applied fivefold. Removing five symmetry-related wedges from a cylindrical shell and stitching the exposed $(111)$ facets produces five partial wedge disclinations, as illustrated in Fig.~\ref{fig:zbar_geometric_construction}(e). As their apex lines are moved toward the outer boundary, the shell evolves into the pentagonal NW shown in Fig.~\ref{fig:zbar_geometric_construction}(f). Along the global $M_y$ mirror plane, the TP-edge and opposite face-midline channels retain the same relative chiralities as the channels at A and B in the single-TP shell. The remaining four pairs follow by successive $C_5$ rotations.
	
	We stress that this construction is restricted to the vicinity of $\Zbar$ and should not be interpreted as an adiabatic deformation of the complete NW spectrum. Near $\Gammabar$, the axis of a partial wedge disclination can bind a helical pair whose presence depends on whether the emerging TP is cationic or anionic. For a cationic TP, moving this axis to the outer boundary produces the TP-edge Dirac mode discussed in the main text. The construction also assumes that the buried disclination line does not carry an additional helical pair near $\Zbar$. Such a pair would affect the spectrum when the TP reaches the outer surface. The gapped spectra of the single-TP shells in Fig.~\ref{fig:TP_slab_spectra}(a,b) and of the $C_5$-resolved NWs in Fig.~\ref{fig:c5_resolved_nw}(a,b) show no such effect near $\Zbar$ for either TP sublattice type. Figure~\ref{fig:zbar_geometric_construction} should therefore be regarded as a geometric conceptualization of the opposite velocities in the single-TP shell and NW geometries, rather than as an independent proof. The direct evidence for the resulting gap is provided by the numerical spectra and their consistency with the low-energy model.
	
	The geometric construction can be translated directly into the doubled low-energy theory. In the cylindrical-shell limit, the two domain walls are related by a rotation through $\pi$ that exchanges the TP-edge and face-midline sectors while rotating their spin polarizations. In the self-conjugate rotational sector $\lambda_0=-1$, it may be represented as
	\begin{equation}
		\widetilde C_2^{\{0\}}=
		\begin{pmatrix}
			0&-is_z\\
			-is_z&0
		\end{pmatrix}.
		\label{eq:zbar_auxiliary_c2}
	\end{equation}
	For $\nu=0$, the constraints in Eq.~\eqref{eq:10_constraints} already imply
	\begin{equation}
		\Delta_{h,0}=
		\Delta_{f,0}=
		t'_0=
		u'_0=0.
		\label{eq:zbar_m0_constraints}
	\end{equation}
	Requiring the doubled Hamiltonian $\widetilde H^{\{-1\}}$ to be invariant under the auxiliary operation in Eq.~\eqref{eq:zbar_auxiliary_c2} further gives
	\begin{equation}
		\epsilon_{h,0}=\epsilon_{f,0},
		\qquad
		v_{h,0}=-v_{f,0},
		\qquad
		u_0=0.
		\label{eq:zbar_c2_constraints}
	\end{equation}
	The opposite velocities on the two domain walls in the cylindrical shell thus follow algebraically from the same $\pi$ rotation that reverses the global mirror assignment in the geometric construction. Since $C_2$ is not a symmetry of the pentagonal NW, the equality of the velocity magnitudes need not persist away from the cylindrical-shell limit.
	
	\subsection{Realistic-model confirmation}
	
\begin{figure}
		\centering
		\includegraphics[scale=1]{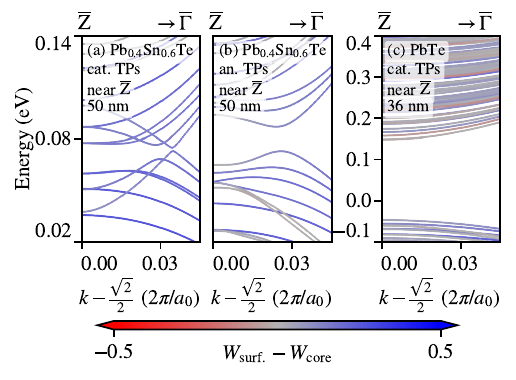}
		\caption{Band structures near $\Zbar$ obtained with the $sp^3d^5$ model for the same NW systems as in Fig.~\ref{fig:lent_bs}. Panels (a) and (b) show the spectra of 50-nm-thick $\mathrm{Pb}_{0.4}\mathrm{Sn}_{0.6}\mathrm{Te}$ NWs with cationic and anionic TPs, respectively, while panel (c) shows the topologically trivial 36-nm-thick PbTe NW with cationic TPs. Red and blue color intensities indicate wave-function weight near the core and the outer surface, respectively.}
		\label{fig:zbar_realistic}
	\end{figure}

	Finally, we examine the spectra near $\Zbar$ in the material-specific $sp^3d^5$ model. The calculations are performed for the same pentagonal NW systems considered in the main text in Fig.~\ref{fig:lent_bs}. Figure~\ref{fig:zbar_realistic} shows the corresponding spectra for 50-nm-thick Pb$_{0.4}$Sn$_{0.6}$Te NWs with cationic and anionic TPs and for the PbTe control system. Both Pb$_{0.4}$Sn$_{0.6}$Te spectra remain gapped at $\Zbar$, although their detailed low-energy subband structures depend strongly on the TP sublattice type. 
	
	\section{Structural stability of pentagonal NWs}
	\label{app:str_stability}
	
	We employ density functional theory (DFT) calculations to examine the binding energies and structural relaxation of pentagonal SnTe NWs. The calculations were performed using the Vienna \textit{Ab initio} Simulation Package (VASP)~\cite{Kresse_1993,Kresse_1996}, with the same computational parameters as in Ref.~\cite{hussain2024pentagonal}. All atomic structures were relaxed without spin--orbit coupling (SOC), which is expected to affect the equilibrium geometries only weakly. The relaxations were unconstrained except for the symmetrized stoichiometric-core configuration, in which the two core atoms were constrained to remain at their initial positions on the NW axis. SOC is included in the electronic band-structure calculations discussed below.
	
	First, we compare bulk SnTe and cubic NWs with square cross sections grown along the $[001]$ direction with the pentagonal NWs considered in this work. The latter include several core configurations. In the non-stoichiometric occupied-core configuration used in the main text, the five TPs meet at a single uncompensated atomic column composed of cations for cationic TPs and anions for anionic TPs. We also consider the stoichiometric core-chain configuration introduced in Ref.~\cite{hussain2024pentagonal}, containing one cation and one anion per axial unit cell, as well as a hollow-core configuration obtained by removing the central column from the non-stoichiometric NW.
	
	Following Ref.~\cite{hussain2024pentagonal}, we define the binding energy per atom as
	\begin{equation}
		E_B=
		\frac{
			n_{\mathrm{Sn}}E_{\mathrm{Sn}}
			+n_{\mathrm{Te}}E_{\mathrm{Te}}
			-E_{\mathrm{cell}}
		}{
			n_{\mathrm{Sn}}+n_{\mathrm{Te}}
		},
		\label{eq:binding_energy}
	\end{equation}
	where $n_{\mathrm{Sn}}$ and $n_{\mathrm{Te}}$ are the numbers of Sn and Te atoms in the periodic unit cell, $E_{\mathrm{Sn}}$ and $E_{\mathrm{Te}}$ are the total energies of the corresponding isolated atoms, and $E_{\mathrm{cell}}$ is the total energy of the periodic atomic structure. A larger value of $E_B$ therefore corresponds to stronger binding per atom.
	
\begin{figure}
		\centering
		\includegraphics[width=0.4\textwidth]{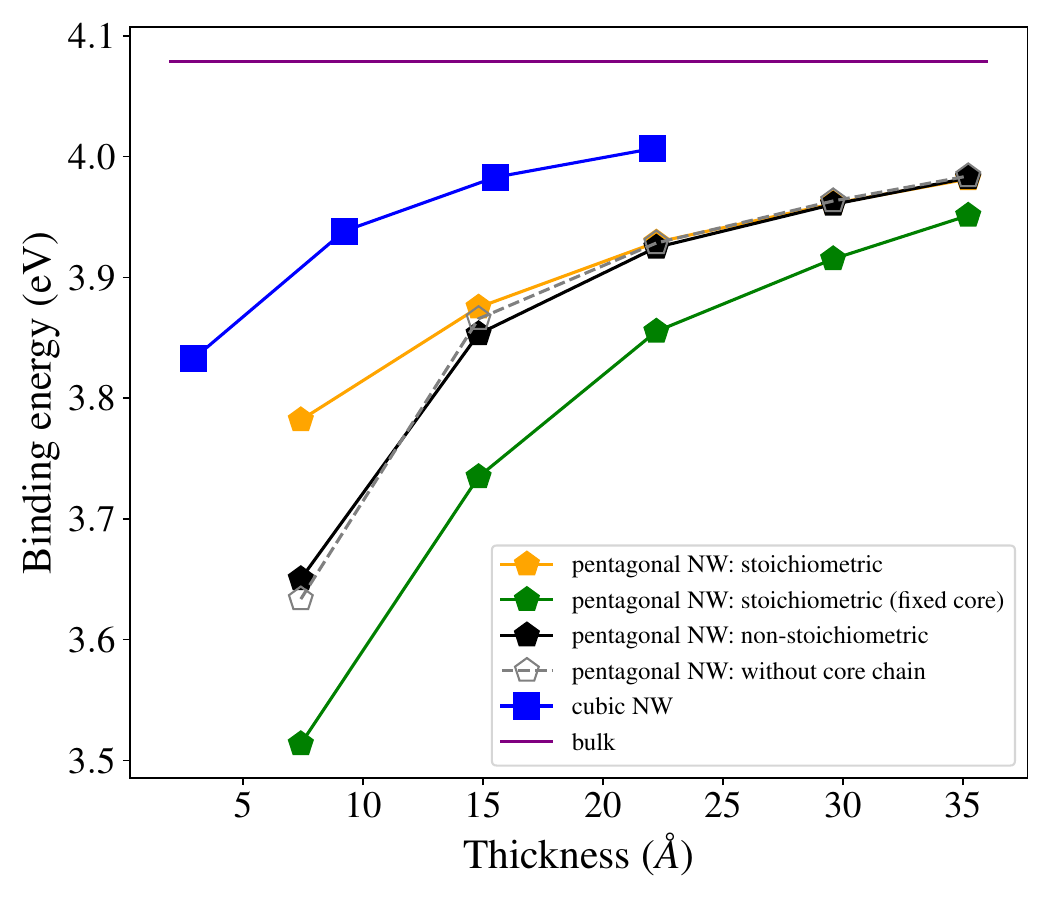}
		\caption{Thickness dependence of the binding energy of SnTe NWs calculated without SOC. The horizontal purple line marks the bulk SnTe value, while the blue curve corresponds to cubic NWs with square cross sections. The black, orange, green, and dashed gray curves represent pentagonal NWs with cationic TPs and non-stoichiometric occupied, relaxed stoichiometric, symmetrized stoichiometric, and hollow cores, respectively.}
		\label{fig:binding}
	\end{figure}

	The calculated binding energies are shown in Fig.~\ref{fig:binding}. In agreement with Ref.~\cite{hussain2024pentagonal}, the square-cross-section NWs are more strongly bound than the pentagonal NWs, indicating that the pentagonal morphology is metastable relative to the square geometry. Within each structural family, $E_B$ increases with wire thickness as the relative contributions of the outer surfaces and the axial star disclination decrease. The effect of the axial period on $E_B$ was checked using doubled unit cells for two atomic configurations after unconstrained structural relaxation: the non-stoichiometric occupied-core NW and the stoichiometric-core NW, both with cationic TPs. The resulting changes in the binding energies were negligible, and the shortest axial unit cell is therefore used in the remaining calculations.
	
\begin{figure*}
		\centering
		\includegraphics[scale=1]{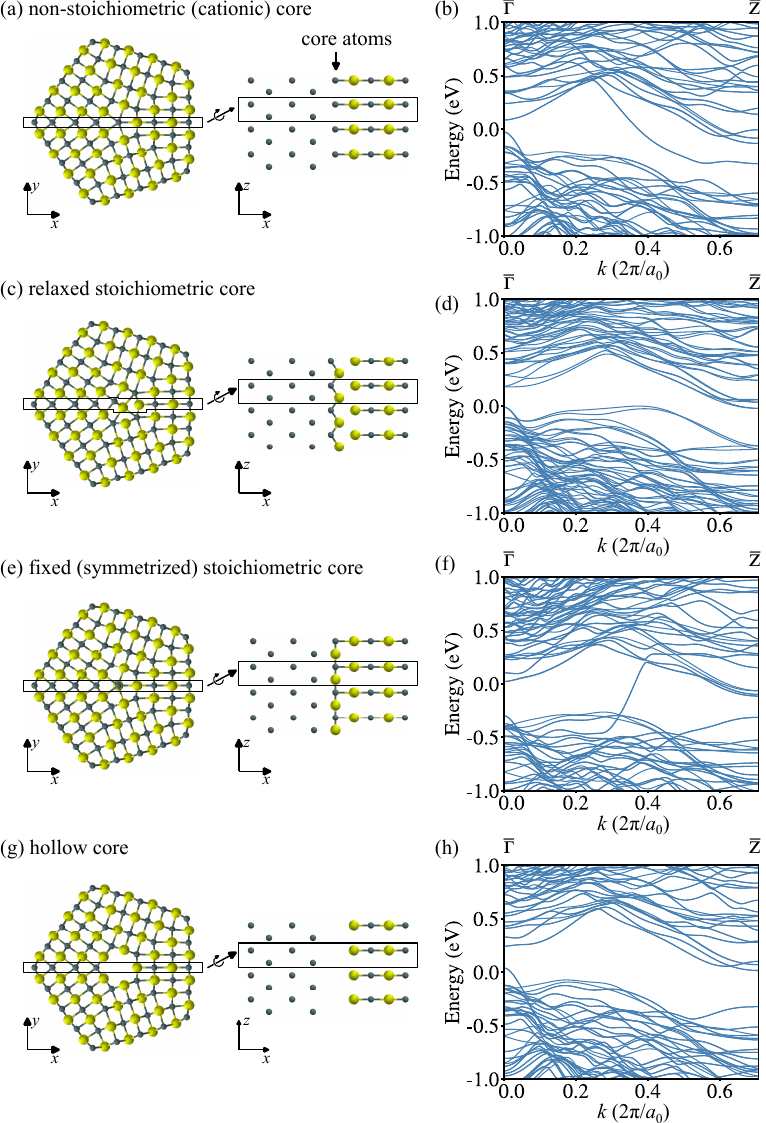}
		\caption{Atomic structures and electronic band structures of four-ring pentagonal SnTe NWs with cationic TPs. Panels (a), (c), (e), and (g) show the atomic structures of the non-stoichiometric occupied-core, relaxed stoichiometric-core, symmetrized stoichiometric-core, and hollow-core NWs, respectively. Each panel includes a cross section perpendicular to the NW axis and a longitudinal section along the central plane. Panels (b), (d), (f), and (h) show the corresponding electronic band structures along the $\Gammabar$--$\Zbar$ line. SOC is included in the band-structure calculations.}
		\label{fig:bs_dft}
	\end{figure*}

	The fully relaxed pentagonal NWs with non-stoichiometric and stoichiometric cores have similar binding energies at thicknesses above $1.5$~nm, as indicated by the black and orange curves in Fig.~\ref{fig:binding}. The stoichiometric atomic configuration is obtained from the high-symmetry core chain introduced in Ref.~\cite{hussain2024pentagonal}, in which a cation and an anion are aligned along the NW axis and initially separated by $\sqrt{2}a_0/4$. This separation is substantially smaller than the bulk nearest-neighbor distance $a_0/2$ and is therefore energetically unfavorable. Under unconstrained relaxation, the two core atoms move apart and the core loses its ideal $C_5$ symmetry, as shown in the transverse and longitudinal views in Fig.~\ref{fig:bs_dft}(c), demonstrating that the high-symmetry atomic configuration is not a stable local minimum. The energetic cost of retaining the symmetrized geometry is evaluated by constraining the two core atoms to their initial axial positions. The resulting binding energies, shown by the green curve in Fig.~\ref{fig:binding}, are lower than those of the fully relaxed atomic structures. The hollow-core configuration has a binding energy close to that of the non-stoichiometric occupied-core configuration, as shown by the dashed gray curve.
	
	We next calculate the electronic band structures of four-ring pentagonal SnTe NWs with cationic TPs. The atomic structures are shown in Figs.~\ref{fig:bs_dft}(a,c,e,g) through cross sections perpendicular to the NW axis and longitudinal sections along the central plane. The corresponding spectra along the $\Gammabar$--$\Zbar$ line are shown in Figs.~\ref{fig:bs_dft}(b,d,f,h). The spectra of the fully relaxed non-stoichiometric, fully relaxed stoichiometric, and hollow-core NWs are gapped, as shown in panels (b), (d), and (h), respectively. By contrast, the symmetrized stoichiometric-core NW, whose atomic structure is shown in panel (e), exhibits an isolated band connecting the valence and conduction subbands in panel (f), consistently with Ref.~\cite{hussain2024pentagonal}. This metallic band is tied to the constrained high-symmetry core-chain configuration and disappears under unconstrained structural relaxation.
	
	Because our DFT calculations are restricted to narrow four-ring NWs, the core and outer-surface states overlap and hybridize strongly. Their gapped spectra therefore do not contradict the spatially separated core and surface Dirac modes found in the larger-radius NWs studied in the main text.
	
\begin{figure}
		\centering
		\includegraphics[width=0.4\textwidth]{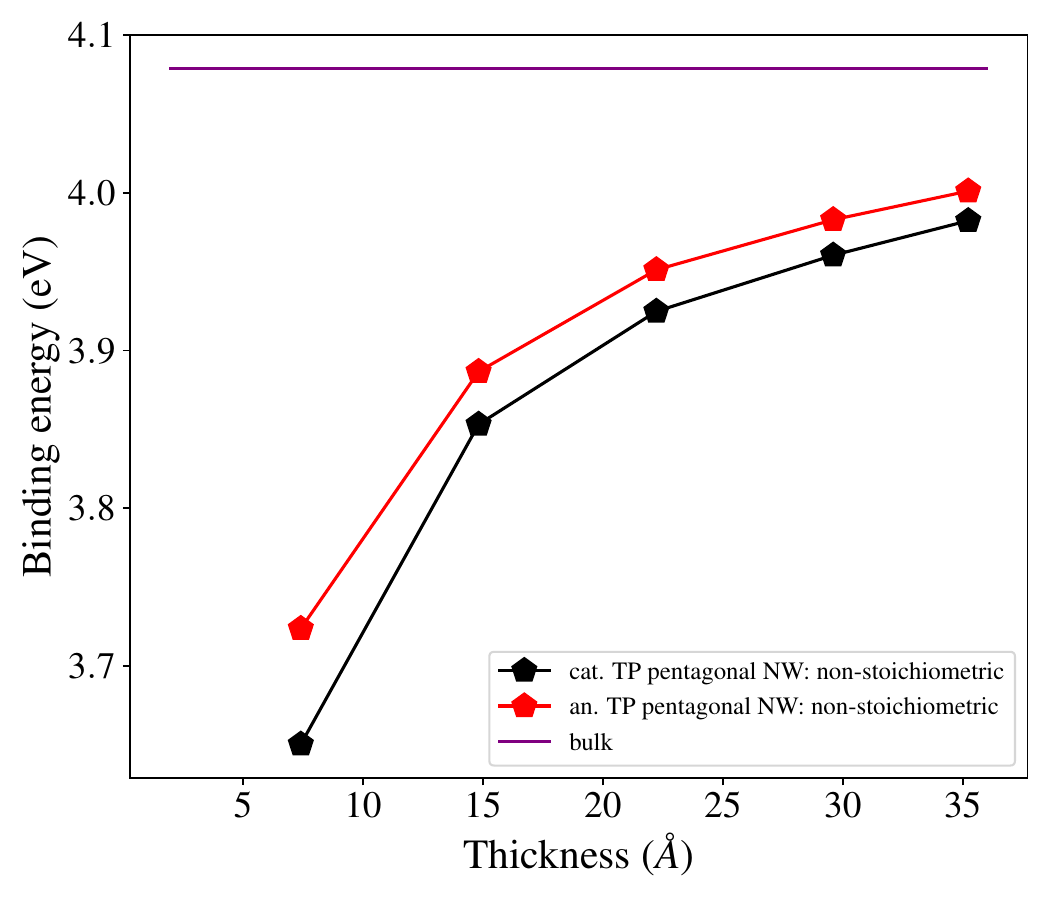}
		\caption{Binding energies of non-stoichiometric pentagonal SnTe NWs with cationic TPs (black) and anionic TPs (red), calculated without SOC. The horizontal purple line marks the bulk SnTe value.}
		\label{fig:binding_an_cat}
	\end{figure}

	The relative stability of the two TP sublattice types is shown in Fig.~\ref{fig:binding_an_cat}. Within the investigated thickness range, the non-stoichiometric NWs with anionic TPs are more strongly bound than their cationic-TP counterparts.
	
	\newpage
	\bibliography{ref}
	
\end{document}